\documentclass[12pt]{article}
\usepackage{graphicx,psfrag}
\usepackage{bm}
\begin{document}

%\preprint{APS/000}

\title{Coexisting solutions and their neighbourhood
in the dynamical system describing second-order optical processes }

\author{I. \'{S}liwa \footnote{E-mail: izasliwa@amu.edu.pl},
P. Szlachetka \footnote{E-mail: przems@amu.edu.pl} and K. Grygiel
\footnote{E-mail: grygielk@amu.edu.pl} \\
\em{Nonlinear Optics Division, Institute of Physics,}\\
\em{A. Mickiewicz University, ul. Umultowska 85,}\\
\em{PL 61-614, Pozna\'{n}, Poland } }

\date{\today}
\maketitle

\begin{abstract}
 Coexisting periodic solutions of a dynamical system describing nonlinear
optical processes of the second-order are studied. The analytical results concern both the
 simplified autonomous model and the extended nonautonomous model, including the
pump and damping mechanism. The nonlinearity  in the coexisting solutions of the
autonomous system is  in concealed frequencies depending on the initial conditions.
 In the solutions of the nonautonomous system the nonlinearity is convoluted in
amplitudes. The neighbourhood of periodic solutions is studied numerically, mainly in
 phase portraits. As a result of disturbance, for example detuning, the periodic solutions are
 shown to escape to other states, periodic,  quasiperiodic (beats)
or chaotic. The chaotic behavior is
indicated by the Lypunov exponents. We also investigate selected aspects of
synchronization (unidirectional or mutual) of two identical systems being in two
different coexisting states. The effects of quenching of oscillations are shown. In the
autonomous system
the quenching is caused by a change in frequency, whereas  in the nonautonomous one
by a change in amplitude. The quenching seems very promising for design of some advanced signal
processing.
\end{abstract}

%\pacs{ 05.45.Xt, 42.65.Sf}
\newpage
\section{Introduction}
Nonlinear systems are usually characterized by two or more coexisting states, corresponding to
 the same values of parameters. As for the first
 time found by Poincare', a  periodic solution disappears (or appears)
 by couples in the form of real roots of an algebraic equation.
 Today it is well known that both chaotic and periodic states can exist within standard
 nonlinear models like Duffing oscillator
 \cite{Duffing}, Lorenz and Rossler systems or Henon map \cite{Henon}. The dynamical coexistence
 is frequently referred to as a generalized multistability \cite{Arecchi, arecchi87}. Multistable
 behavior appears also in nonlinear optics \cite{Brun}, electronic circuits \cite{Maurer}
 mechanical systems \cite{Thompson} and neurons nets \cite{Goldbeter, Foss}.
 There is no universal (global) method for detecting multistability
 of a dynamical system. Usually, we do this numerically, for example, looking for
 different basins of attraction in a system with selected parameters. Obviously,
 the numerical approach is necessary if we try to find coexisting chaotic states.
 In some cases, however, analytical form of coexisting states (solutions) is also
 possible to find, provided that the states are regular (periodic). Frequently,
 purely numerical investigation does not deliver sufficient information about the
 intricate nature of coexistence. Therefore, attempts at finding some analytical
 results, if it is possible, are always physically valuable. We try to follow such
 an approach in this paper.
 The aim of this study is to find a class of coexisting periodic solutions in a well
 known nonlinear model describing the nonlinear optical processes of the second order
 and to investigate the behaviour of these analytic solutions in response to a disturbance
 of the source of periodicity. The dynamical model is considered in a simplified version,
 that is
 autonomous and  nonautunomous ones. The dynamics of the autonomous and nonautonomous
  systems are compared within the phase space. The coexisting periodic solutions
  in the nonautonomous model are shown to be controlled by changing the system's parameters,
  which leads to  transitions from one state to another. The coexisting states
 can escape  to chaotic or nonchaotic (periodic) states. The type of the final state
 is deducted from the Lypunov exponents. Finally, the behaviour of two identical
 dynamical systems being in two different coexisting states on a linear interaction between
 the systems  turned on, is studied. In some cases  one or two of the coexisting
 states can be quenched.  The quenching effects are shown to be controllable by
  the parameters of the system.

\section{Equations of motion}

%  \begin{eqnarray}
%\label{e}
%H=\omega a^{*}a +2\omega b^{*}b -\frac{1}{2}i\epsilon(a^{2}b^{*}-ba^{*2})
%\nonumber\\
% +i F_{1}(a^{*}e^{-i\Omega_{1}t} -ae^{+i\Omega_{1}t})
%\nonumber\\
 %+i F_{2}(b^{*}e^{-i2\Omega_{2}t} -be^{+i2\Omega_{2}t}).
%\end{eqnarray}

Let us consider a nonautonomous dynamical system governed by the
following set of equations \cite{Drummond,
Mandel, Savage, Gao}:
\begin{eqnarray}
\label{e1}
\frac{da}{dt}&=&-i\omega a -\gamma_{1}a +\epsilon a^{*}b +F_{1}e^{-i\Omega_{1}t}\,,\\
\label{e2}
\frac{db}{dt}&=&-i2\omega b -\gamma_{2}b -\frac{1}{2}\epsilon a^{2}+F_{2}e^{-i\Omega_{2}t}\,.
\end{eqnarray}
Physically, the equations describe an interaction between two optical modes of the frequencies
$\omega$ and $2\omega$. The complex dynamical variables $a$ and $b$ are the
amplitudes of the fundamental
 and second-harmonics modes, respectively. The interaction takes place
 via a nonlinear crystal placed within a
 Fabry-Pe\'rot interferometr. The quantity $\epsilon$ is a nonlinear coupling coefficient,
 whose value is proportional to the second-order nonlinear susceptibility. The parameters
 $\gamma_{1}$ and $\gamma_{2}$ are the damping constants of the fundamental and
  second-harmonics modes, respectively. Moreover, the system is pumped by two external fields
$F_{1}e^{-i\Omega_{1} t}$ and $F_{2}e^{-i\Omega_{2} t}$, where $F_{1}$ and $F_{2}$ are electric
 field amplitudes at the frequencies $\Omega_{1}$ and $\Omega_{2}$. respectively.
Henceforth, all the parameters, that is $\omega$, $\epsilon$, $F_{1,2}$, and  $\Omega_{1,2}$
 are taken to be real as in \cite{Mandel}.
\\
To visualize the dynamics of the system(\ref{e1})--(\ref{e2}) the
four-dimensional space
\\($\mathit{Re}\,a,\mathit{Re}\,b,\mathit{Im}\,a,\mathit{Im}\,b$)\\
 is required; but, as it is impossibile, we carry out the visualisation
 for its two dimensional sections.
 The system(\ref{e1})--(\ref{e2})
  does not belong to the class of integrable systems and
usually it is studied numerically. However, in special cases, analytical solutions of
(\ref{e1})--(\ref{e2}) are also possible. Below, we consider a
class of coexisting periodic solutions of the system
and  qualify the kind of motion in their neighbourhood.

\subsection{Autonomous case}
\begin{figure}
\includegraphics[width=4cm,height=4cm,angle=0]{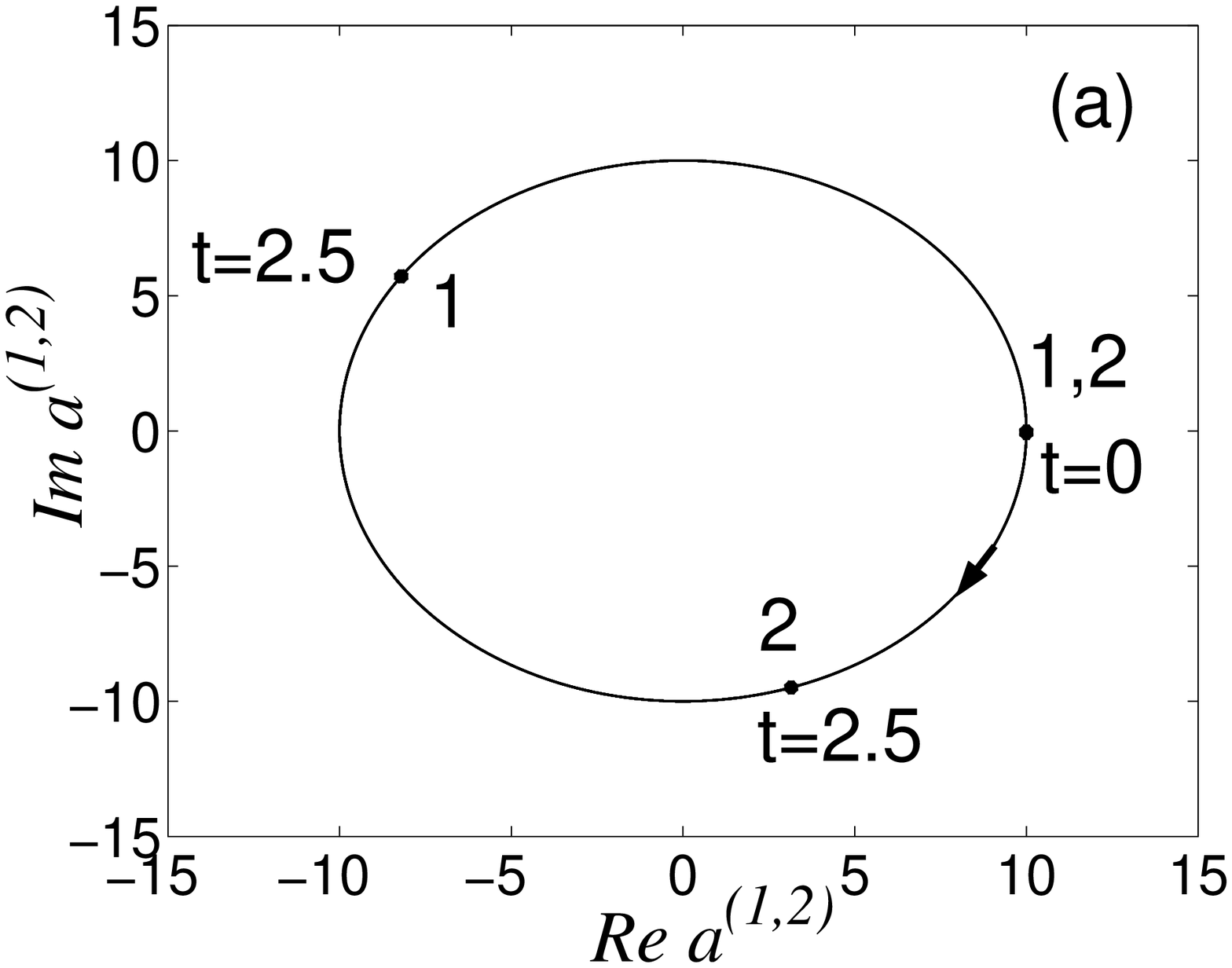}
\includegraphics[width=4cm,height=4cm,angle=0]{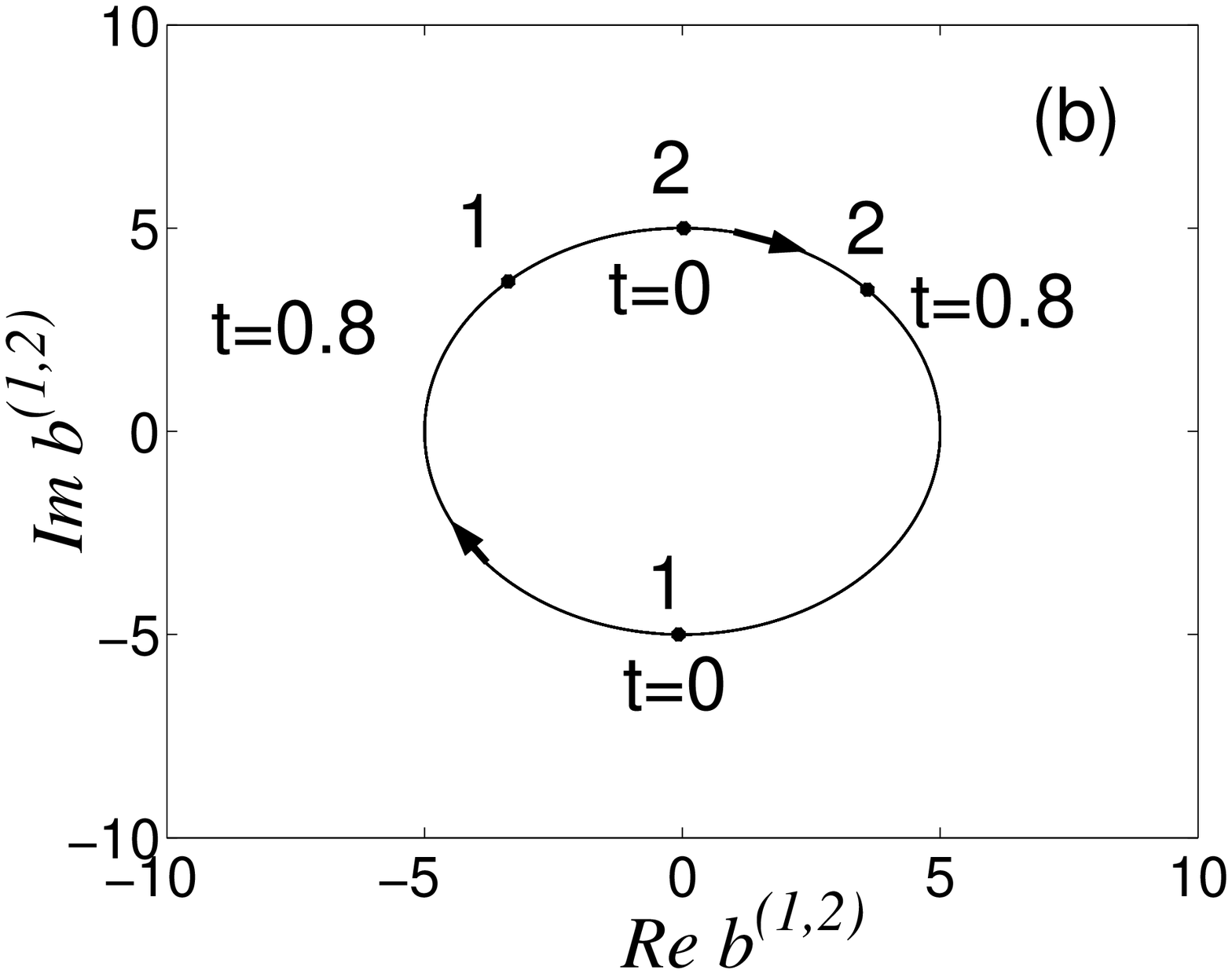}
\includegraphics[width=4cm,height=4cm,angle=0]{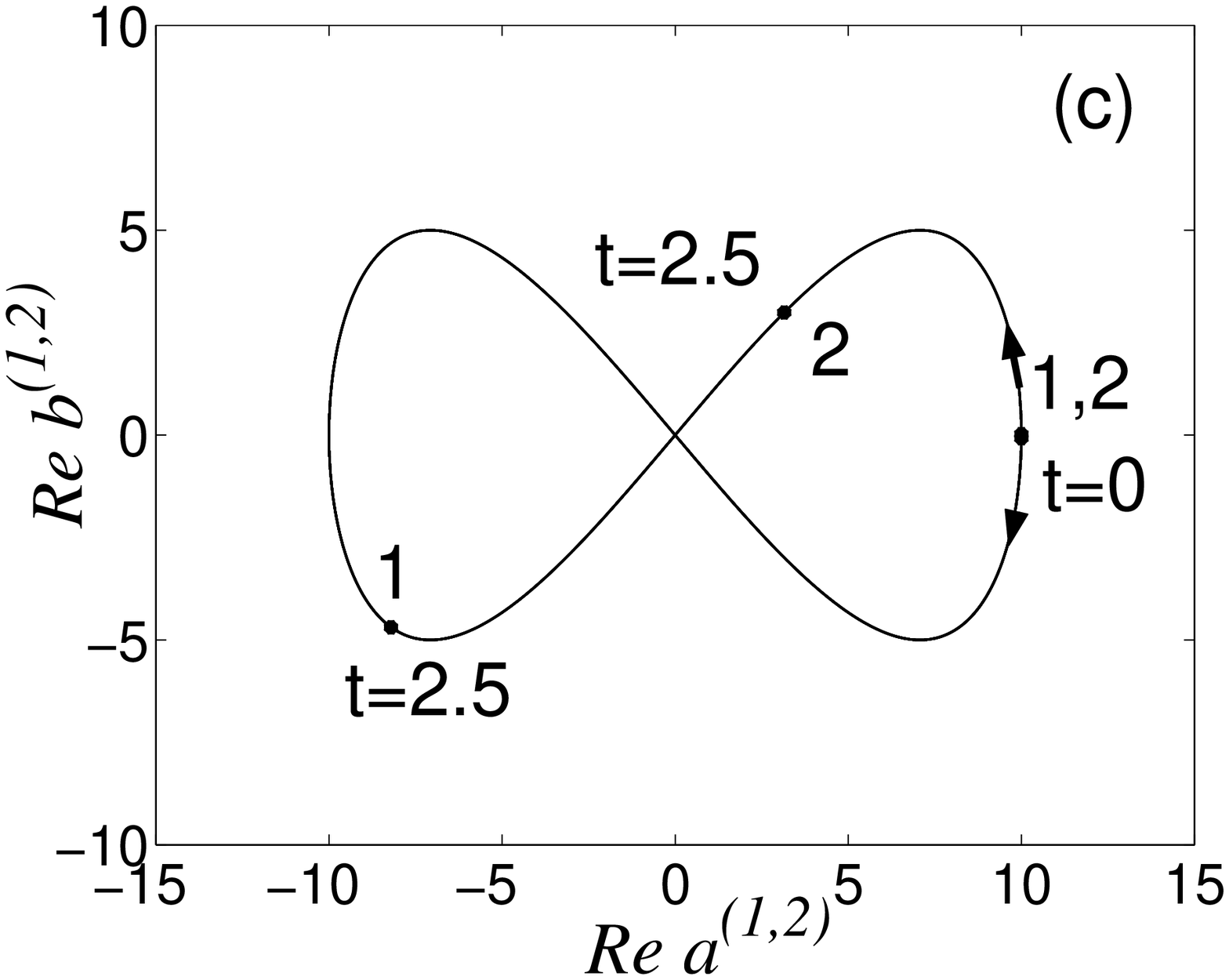}
\includegraphics[width=4cm,height=4cm,angle=0]{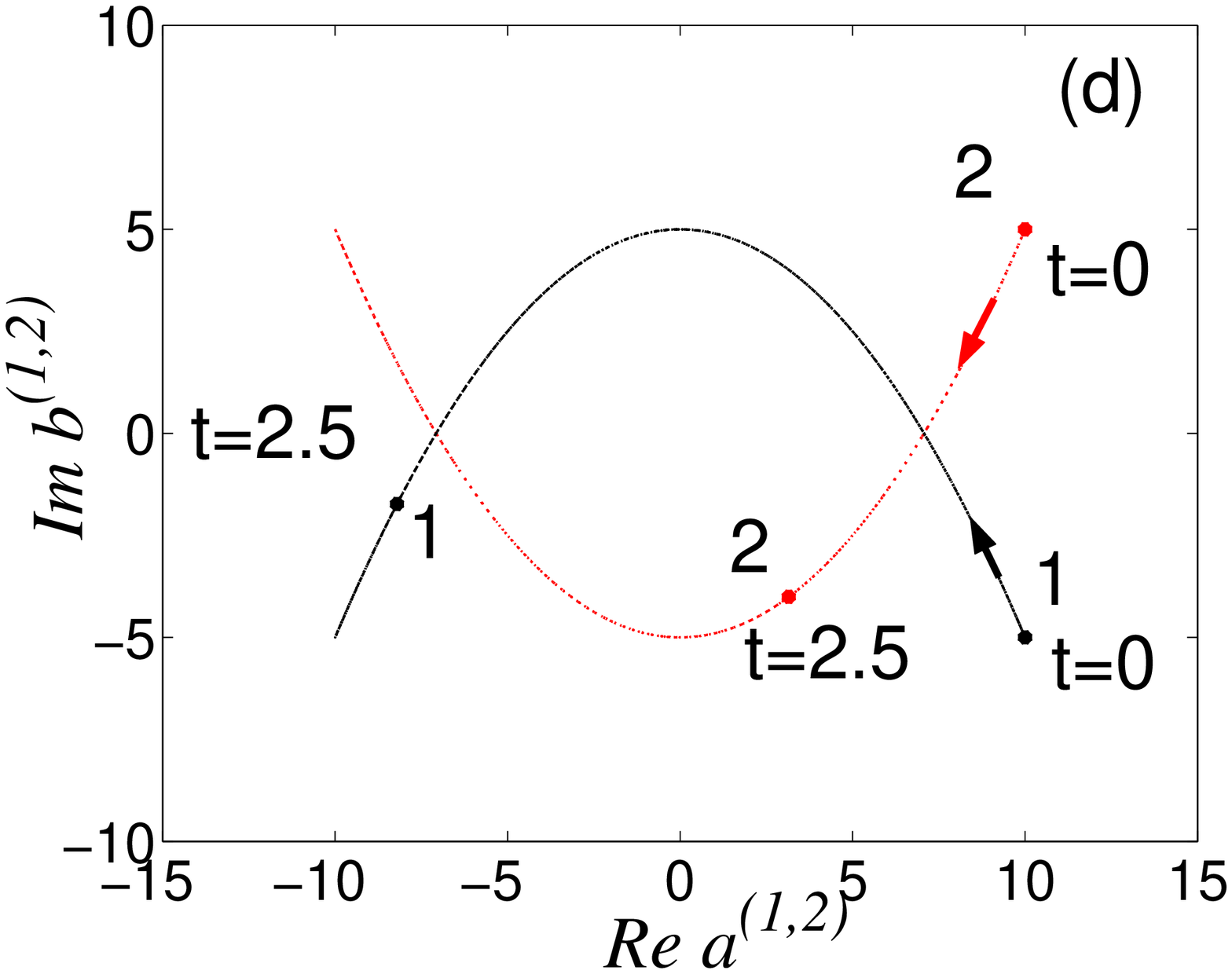}
\includegraphics[width=4cm,height=4cm,angle=0]{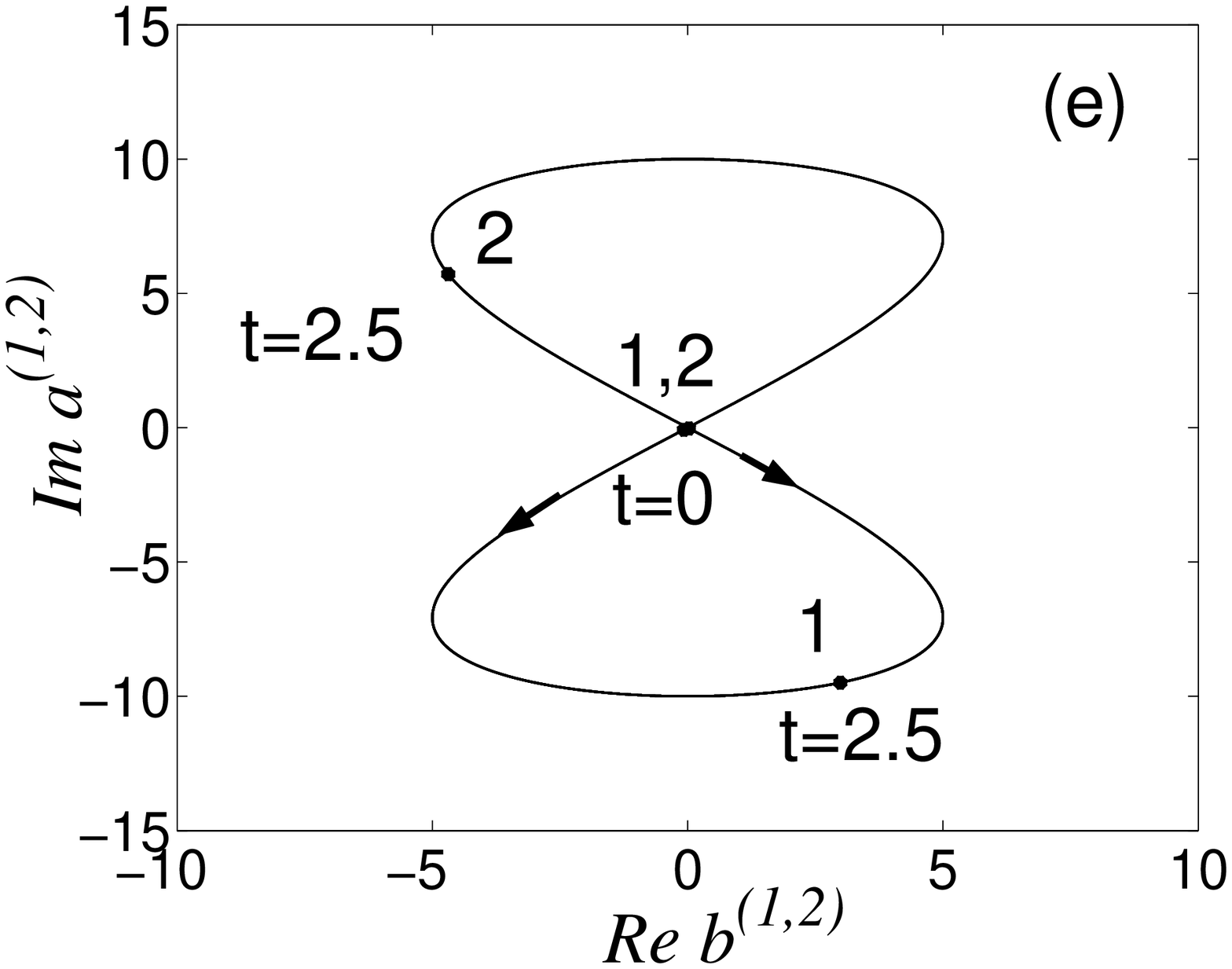}
\includegraphics[width=4cm,height=4cm,angle=0]{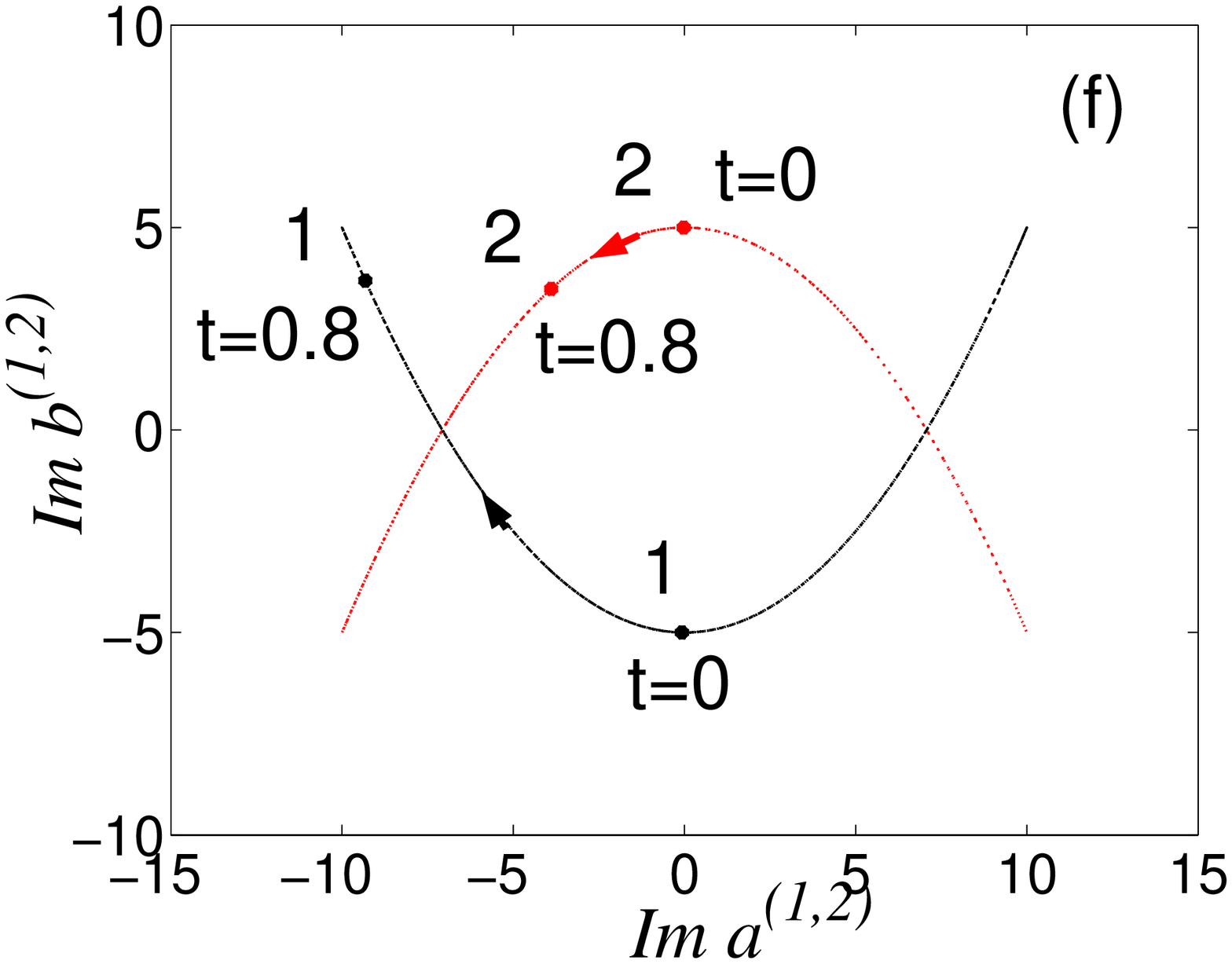}
\caption{Phase-diagrams of the coexisting solutions (\ref{e3})--(\ref{e4})
 and (\ref{e5})-- (\ref{e6}) in black and red, respectively. The parameters are:
  $\mathit{Re}\,\alpha=10$, $\mathit{Im}\,\alpha=0$, $\omega=1$ and $\epsilon=0.1$.
  The positions
 of the phase points 1 and 2  are marked at the initial time $t=0$ and at
  the later times $t=0.8$
 or $t=2.5$.}
\label{fig.1}
\end{figure}

Let us first consider the problem of periodic orbits in the simplest (conservative) version
of the  system(\ref{e1})--(\ref{e2}), that is
\begin{eqnarray}
\label{e1a}
\frac{da}{dt}&=&-i\omega a  +\epsilon a^{*}b \,,\\
\label{e2a}
\frac{db}{dt}&=&-i2\omega b  -\frac{1}{2}\epsilon a^{2}\,.
\end{eqnarray}
The equations of motion (\ref{e1a})--(\ref{e2a}) were used for the first
time by Bloembergen
 to describe  second-harmonic generation of light \cite{bloembergen,Drobny,perina}.
The above system has two coexisting periodic solutions
(the details may be found in Appendix A). The first
\begin{eqnarray}
\label{e3}
a^{(1)}(t)&=&\alpha e^{-i(\omega+ \frac{1}{2}\epsilon\sqrt{\alpha^{*}\alpha}\,)t}\,,\\
\label{e4}
b^{(1)}(t)&=&-\frac{i}{2}\sqrt{\frac{\alpha^{3}}{\alpha^{*}}}e^{-i2(\omega+
\frac{1}{2}\epsilon \sqrt{\alpha^{*}\alpha}\,)t}
\end{eqnarray}
and the second
\begin{eqnarray}
\label{e5}
a^{(2)}(t)&=&\alpha e^{-i(\omega-
\frac{1}{2}\epsilon\sqrt{\alpha^{*}\alpha}\,)t}\,,\\
\label{e6}
b^{(2)}(t)&=&+\frac{i}{2}\sqrt{\frac{\alpha^{3}}{\alpha^{*}}}e^{-i2(\omega-
\frac{1}{2}\epsilon \sqrt{\alpha^{*}\alpha}\,)t}\,.
\end{eqnarray}
The coexisting first harmonics (\ref{e3}) and (\ref{e5}) have identical amplitudes
$\alpha$  and different frequencies $\omega+
0.5\epsilon \sqrt{\alpha^{*}\alpha}$ and $\omega-
0.5\epsilon \sqrt{\alpha^{*}\alpha}$,  being functions of the
 initial condition $\alpha$.
 The  second harmonics (\ref{e4}) and
(\ref{e6})  have the same amplitudes
$\frac{i}{2}\sqrt{\frac{\alpha^{3}}{\alpha^{*}}}$ and different
  coexisting frequencies $2\omega+
\epsilon \sqrt{\alpha^{*}\alpha}$ and $2\omega-
\epsilon \sqrt{\alpha^{*}\alpha}$. Additionally, the function $b^{(1)}$ is of the sign opposite
 to that of $b^{(2)}$.\\
Let us note that the functions (\ref{e5}) -- (\ref{e6}) can be constant in time
(the period $T=\infty$) if $ \omega=
\frac{1}{2}\epsilon \sqrt{\alpha^{*}\alpha}$. Physically, it means that the vibrations
are quenched.
The fact that a frequency (period) depends on
  amplitude is well known in the theory of autonomous systems \cite{minorsky}.
  A variation of the period with amplitude is well known, for
  example, in the case of a pendulum for larger deviations.\\
In the
four-dimensional phase space $(\mathit{Re}\,a,\mathit{Re}\,b,\mathit{Im}\,a,\mathit{Im}\,b)$
the solutions (\ref{e3})--(\ref{e4}) and (\ref{e5})--(\ref{e6})
generate two coexisting hyper-surfaces. The geometrical relationship between them are illustrated
in six two-dimensional phase diagrams
(Fig.\ref{fig.1}), where the coexisting solutions create
 simple Lissajous-like curves. Some of them are identical (degenerate), as readily seen,
 for example, in Fig.1a.
Both phase points $1$ and $2$ start together from the same position $(t=0)$,
 rotate in the same direction and   draw the identical orbits
$(\mathit{Re}\,a^{(1)})^{2}+(\mathit{Im}\,a^{(1)})^{2}=r^{2}$ and
$(\mathit{Re}\,a^{(2)})^{2}+(\mathit{Im}\, a^{(2)})^{2}=r^{2}$, where $r^{2}=\alpha^{*}\alpha$.
The only difference is, that the phase point $1$  draws the circle at the frequency
$\omega+
\frac{1}{2}\epsilon \sqrt{\alpha^{*}\alpha}$, whereas the phase point  2 at
the frequency$\omega-
\frac{1}{2}\epsilon \sqrt{\alpha^{*}\alpha}$. The other Lissajous-like trajectories
are presented in Fig.1b - Fig.1f. As seen  in Figs.(a,c,e),  points 1 and 2 start from
 the same position, whereas in Figs.( b,d,f) from two different ones.\\
 Generally, if we start  from any
point lying on a periodic trajectory  of the system (\ref{e1a})-- (\ref{e2a})
 we always remain   in the same trajectory. This is frequently called the
 {\em translation}
properties of autonomous systems which means that {\em to a given trajectory
 corresponds an
infinity of motions (solutions) differing from each other by the phase} \cite{minorsky}.
 The question is, however
 the type of trajectories of the system (\ref{e1a}) -- (\ref{e2a})
   when the initial conditions do not lie on periodic orbits. There is no difficulty in
   solving this problem numerically in contradistinction to a general analytical approach.
The numerical analysis shows that all trajectories of the  system
(\ref{e1a}) -- (\ref{e2a}) originating from independent initial
conditions $a(0)=\alpha$ and $b(0)=\beta$ (contrary to the
dependent conditions $a^{(1,2)}(0)=\alpha$ and
$b^{(1,2)}(0)=\pm\frac{i}{2}\sqrt{\frac{\alpha^{3}}{\alpha^{*}}}$
generating periodic orbits) behave in a nonperiodic, mainly quasiperiodc manner.
For example,
when $a(0)=\alpha$ and $b(0)=0$ we get an analytical result known
since the pioneering  work by Bloembergen \cite{bloembergen, perina}:
\begin{eqnarray}
\label{e7} a(t)&=&\alpha\, sech\left( \frac{\sqrt{2}}{2} \alpha
\epsilon t \right)  e^{-i\omega t}
\end{eqnarray}
\begin{eqnarray}
\label{e8}
b(t)=-\frac{\alpha}{\sqrt{2}}\tanh\left(\frac{\sqrt{2}}{2}\alpha
\epsilon t \right) e^{-i2\omega t}
\end{eqnarray}
where  $\alpha =\alpha^{*}$. Obviously, due to the time dependent
amplitudes $\alpha\, sech\left( \frac{\sqrt{2}}{2} \alpha
\epsilon t \right)$ and
$-\frac{\alpha}{\sqrt{2}}\tanh\left(\frac{\sqrt{2}} {2}\alpha
\epsilon t \right)$ the functions $a(t)$ and $b(t)$ are not
formally periodic i.e. $a(t+T)\neq a(t)$ and $b(t+2T)\neq b(t)$,
where $T=\omega/2\pi$. The solutions are frequently referred to
as {\em nearly periodic}, which mearly means
that in the course of time both functions approach a purely periodic motion.\\
 Another question is the change in the harmonic solutions (\ref{e3})--(\ref{e6})
 on inclusion of damping in  the dynamical
system (\ref{e1a})--(\ref{e2a}):
\begin{eqnarray}
\label{e1b}
\frac{da}{dt}&=&-i\omega a -\gamma a +\epsilon a^{*}b \,,\\
\label{e2b}
\frac{db}{dt}&=&-i2\omega b -\gamma b -\frac{1}{2}\epsilon a^{2}\,,
\end{eqnarray}
where, for the sake of simplicity, we have put  $\gamma_{1}=\gamma_{2}=\gamma$.
Using the method \cite{minorsky} proposed by Krylov-Bogolubov
(we do not present the details here)
 we get two coexisting
nonperiodic solutions
$\{a^{(1)}(t),b^{(1)}(t)\}$ and {$\{a^{(2)}(t),b^{(2)}(t)\}$, namely:
\begin{eqnarray}
\label{e9}
a^{(1,2)}(t)&=&\alpha e^{-\gamma t} e^{-i[\omega t \pm
  \frac{\epsilon}{2\gamma}\sqrt{\alpha^{*}\alpha}(1-e^{-\gamma t})]}\,,\\
\label{e10} b^{(1,2)}(t)&=&\mp
\frac{i}{2}\sqrt{\frac{\alpha^{3}}{\alpha^{*}}} e^{-\gamma
t}e^{-i2[\omega
  t\pm
  \frac{\epsilon}{2\gamma }
\sqrt{\alpha^{*}\alpha}(1-e^{-\gamma t})]}\,.
\end{eqnarray}
In the limit  $\gamma\rightarrow 0$, the above functions become periodic
solution (\ref{e3})--(\ref{e6}). Here, in contradistinction to the periodic solutions,
 not only the amplitudes are functions of
the  damping constant $\gamma$ but
 also the phases. For $t \rightarrow \infty$ the coexisting solutions
 $\{a^{(1)}(t),b^{(1)}(t)\}$ as well as {$\{a^{(2)}(t),b^{(2)}(t)\}$  tend to zero. \\
Damping damages degeneration of the periodic orbit in Fig.\ref{fig.1} (a,b,c and e).
 By way of example, it
 is illustrated in Fig.2.
\begin{figure}
\includegraphics[width=5cm,height=5cm,angle=0]{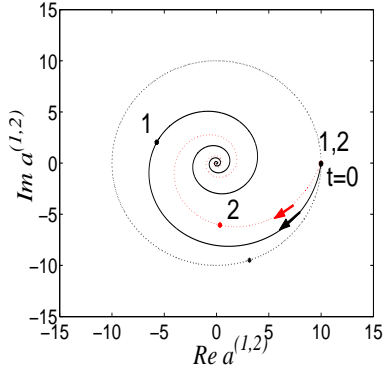}
\caption{Phase diagram of the function (\ref{e9}) for $\gamma=0$ (circle) and
$\gamma=0.2$ (spirals). The circle is identical to that in Fig.1a. The phase points
$1$ and $2$ escape from the joint coexisting orbit due to damping.
The first (black) goes faster then the second (red).
 } \label{fig.2}
\end{figure}
  For $\gamma\neq 0$ the phase points 1 and 2,
 drawing different trajectories ($1$
  goes faster  than $2$), approach the fixed point $(0,0)$,
  being an attractor.  \\
Equations (\ref{e1b})--(\ref{e2b}) may be solved subject to the initial conditions
   $a(0)=\alpha$ and $b(0)=0$ to yield
\begin{eqnarray}
\label{e11} a(t)&=&\alpha e^{-\gamma t}\, sech \left(
\frac{\sqrt{2}}{2\gamma} \alpha \epsilon (1-e^{-\gamma
t})\right)  e^{-i\omega t}\,\\
\label{e12}
 b(t)&=&-\frac{\alpha e^{-\gamma
t}}{\sqrt{2}}\tanh\left(\frac{\sqrt{2}}{2\gamma}\alpha \epsilon
(1-e^{-\gamma t})\right) e^{-i2\omega t}\,.
\end{eqnarray}
The behavior of the above amplitudes and phases is remarkably different from
 that presented by
Eqs.(\ref{e9})--(\ref{e10}). Here, the phases depend neither on the initial
 conditions nor on the
damping constant. The amplitudes are damped in a much more intricate way
 than those in Eqs.
 (\ref{e9})--(\ref{e10}) .\\
 In the phase space, both functions (\ref{e9})--(\ref{e10}) and
 (\ref{e11})--(\ref{e12}) tend to the same attractor being a fixed point.
Obviously, in the limit $\gamma \rightarrow 0$ the functions (\ref{e11})--(\ref{e12}) approach
arbitrarily closely functions (\ref{e7})--(\ref{e8}), respectively.
  Generally, the numerical studies show clearly that in the phase space the
system (\ref{e1b})--(\ref{e2b}) always   tends to a fixed point,
 independently of the initial conditions.

\subsection{Periodic resonance solutions of nonautonomous system }

In order to find a periodic resonance solution
of the system(\ref{e1})-- (\ref{e2})
we look for a solution in the form
\begin{eqnarray}
\label{solll1}
a(t)=X e^{-i\Omega_{1} t}\,\,,\,\,\,\,
b(t)=Y e^{-i\Omega_{2} t}\,,
\end{eqnarray}
where  $\Omega_{1}=\omega$ and $\Omega_{2}=2\omega$ are the resonance conditions. The amplitudes
 $X$ and $Y$ are constant in time.
On inserting  (\ref{solll1}) into (\ref{e1})-- (\ref{e2}) we get two algebraic
 equations (quadratic) in the complex variables
\begin{eqnarray}
\label{e1s}
 -\gamma_{1}X +\epsilon X^{*}Y +F_{1}=0\,,\\
\label{e2s}
 -\gamma_{2}Y -\frac{1}{2}\epsilon X^{2}+F_{2}=0\,.
\end{eqnarray}
Therefore, we look for $X$ and $Y$ as functions of the parameters: $\epsilon$,
$\gamma_{1}$, $\gamma_{2}$,  $F_{1}$ and $F_{2}$. Restricting the number of
 the parameter to three, we get
solutions whose physical context is clear, and the algebraic form is easy
 for numerical investigation.\\
\subsubsection{ The case I,\,\,$F_{1}=0$\,, $F_{2}\neq0$\,,
 $\gamma_{1}\neq0$ and $\gamma_{2}\neq 0$}
This case ($F_{1}=0$) describes the
 subharmonic generation. There are two coexisting solutions
 $\{a^{(1)}(t),b(t)\}$ and $\{a^{(2)}(t),b(t)\}$
\begin{eqnarray}
\label{non1}
a^{(1,2)}(t)&=&\mp \sqrt{
\frac{2F_{2}}{\epsilon}-\frac{2\gamma_{1} \gamma_{2}}{\epsilon^{2}}}\, e^{-i\omega t}\,\,\,\,,F_{2}>\frac{\gamma_{1}\gamma_{2}}{\epsilon}\,,\\
\label{non2}
b(t)&=&\frac{\gamma_{1}}{\epsilon} e^{-i2\omega t}\,.
\end{eqnarray}
which differ only in the phase $a^{(2)}(t)=a^{(1)}(t)e^{-i\pi}$.
 Therefore, the coexistence has a trivial character.
(see Eqs.(3.1, second line) in Ref.\cite{Drummond}).
\subsubsection{ The case II,\,\,$F_{1}\neq 0$\,, $F_{2}=0$, $\gamma_{1}\neq 0$ and $\gamma_{2}\neq 0 $}
 Physically, this case ($F_{2}=0$)
corresponds to the second harmonic generation. There are no coexisting solutions
but only one single solution (see Eqs.(3) and (4) in Ref.\cite{Mandel}):
\begin{eqnarray}
\label{non3}
a(t)&=&(A+B) e^{-i\omega t}\,,\\
\label{non4}
b(t)&=&-\frac{\epsilon}{2\gamma_{2}} (A+B)^{2}  e^{-i2\omega t}\,,\\
\nonumber
\label{non5}
A&=&\sqrt[3]{ \frac{\gamma_{2}F_{1}}{\epsilon^{2}}+\sqrt{\left (\frac{2\gamma_{1}\gamma_{2}}{3\epsilon^{2}}\right)^{3} +\left(\frac{\gamma_{2}F_{1}}{\epsilon^{2}}\right)^{2}}}\,,\\
\nonumber
\label{non6}
B&=&\sqrt[3]{ \frac{\gamma_{2}F_{1}}{\epsilon^{2}}-\sqrt{\left (\frac{2\gamma_{1}\gamma_{2}}{3\epsilon^{2}}\right)^{3} +\left(\frac{\gamma_{2}F_{1}}{\epsilon^{2}}\right)^{2}}}\,.
\end{eqnarray}
\subsubsection{ The case III,\,\,$F_{1}\neq 0$\,, $F_{2}\neq 0$, $\gamma_{1}\neq 0$
 and $\gamma_{2}= 0 $}
Here, both
the subharmonic effect and the second harmonic processes compete with each other.
The assumption $\gamma_{2}=0$  describes the so-called good frequency conversion limit for
 subharmonic generation.
 It is easy to prove that the system (\ref{e1})--(\ref{e2})
 has two coexisting solutions $\{a^{(1)}(t),b^{(1)}(t)\}$
  and $\{a^{(2)}(t),b^{(2)}(t)\}$ given by
\begin{eqnarray}
\label{non7}
a^{(1,2)}(t)&=&\mp \sqrt{\frac{2 F_{2}}{\epsilon} } e^{-i\omega t }\,,\\
\label{non8}
b^{(1,2)}(t)&=&\left(\frac{\gamma_{1}}{\epsilon}\pm\frac{F_{1}}{2F_{2}}\sqrt{\frac{2F_{2}}{\epsilon
}}\,\right) e^{-i2\omega t}\,.
\end{eqnarray}
Physically, for the same values of parameters the system has two periodic states differing
in the vales of amplitude. If
 $\frac{\gamma_{1}}{\epsilon}=\frac{F_{1}}{2F_{2}}\sqrt{\frac{2F_{2}}{\epsilon}}$,
 then  second-harmonic vibrations are quenched  ($b^{(2)}(t)=0$).
 In this case the subharmonic generation is maximal.

Phase diagrams for the coexisting solutions $\{a^{(1)}(t),b^{(1)}(t)\}$ and
$\{a^{(2)}(t),b^{(2)}(t)\}$
are presented in Fig.(\ref{fig.3}).  As seen, only the phase curve for the pair
$(Re\,a^{(1)}(t),Im\, a^{(1)}(t))$
 covers the phase curve for the pair  $(Re\, a^{(2)}(t),Im\, a^{(2)}(t))$.
Here, point $1$ is by $\pi$ out of phase with point $2$.
 The other curves are non-degenerate that is they are separable. This behaviour follows from
 the fact that the functions $b^{(1,2)}(t)$ differ in the values of amplitudes.\\
  There is no geometrical  correspondence between the phase portraits in
 Fig.\ref{fig.1} and Fig.\ref{fig.3}.\\
The main difference is, however, between the autonomous and nonautonomous
phase dynamics.
 The translation properties of  autonomous systems ( sometimes called {\em free}
  phase \cite{pikovsky}) do not
 hold in nonautonomous systems. It means, for example that the phase
 points $1$ and $2$ in
 Fig.\ref{fig.3} (a) follow the circle provided that they start only from the points
  $(a(0)=-10$, $b(0)=0)$ and $(a(0)=+10$, $b(0)=0)$, respectively. If they start
from the other points lying on the circle they escape from it, which does not take
 place in the autonomous case (this problem is considered in detail in Section III).\\
  As  seen from Eqs. (\ref{non7})--(\ref{non8}) the parameter $\epsilon$ governing the
 nonlinearity of the system (\ref{e1})--(\ref{e2}) is felt in the amplitudes, in
 contradistinction to the autonomous case, where $\epsilon$ is felt in the phases
 (see Eqs.(\ref{e9})--(\ref{e10})).
%_________________________________________________________________
 \begin{figure}
\includegraphics[width=4cm,height=4cm,angle=0]{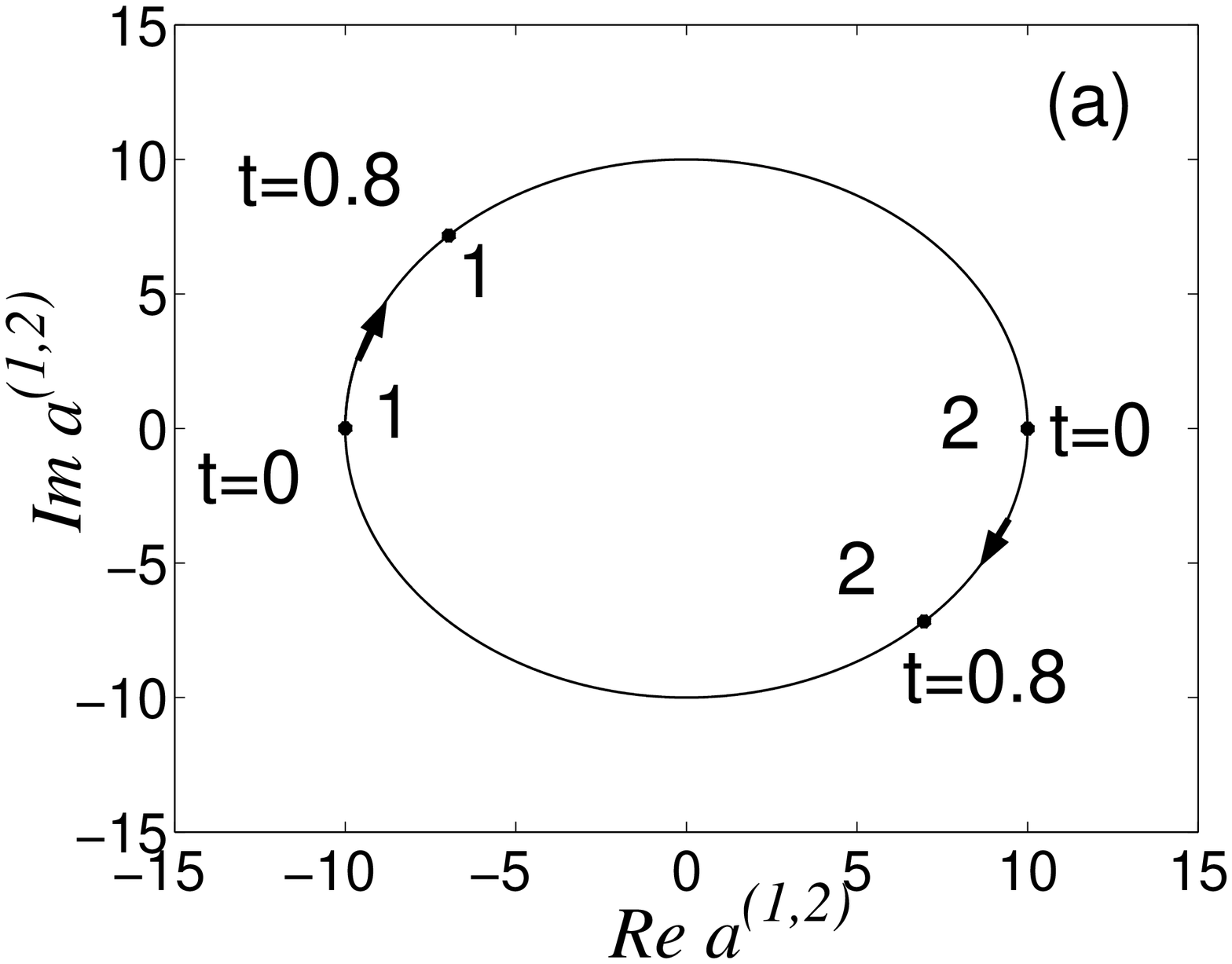}
\includegraphics[width=4cm,height=4cm,angle=0]{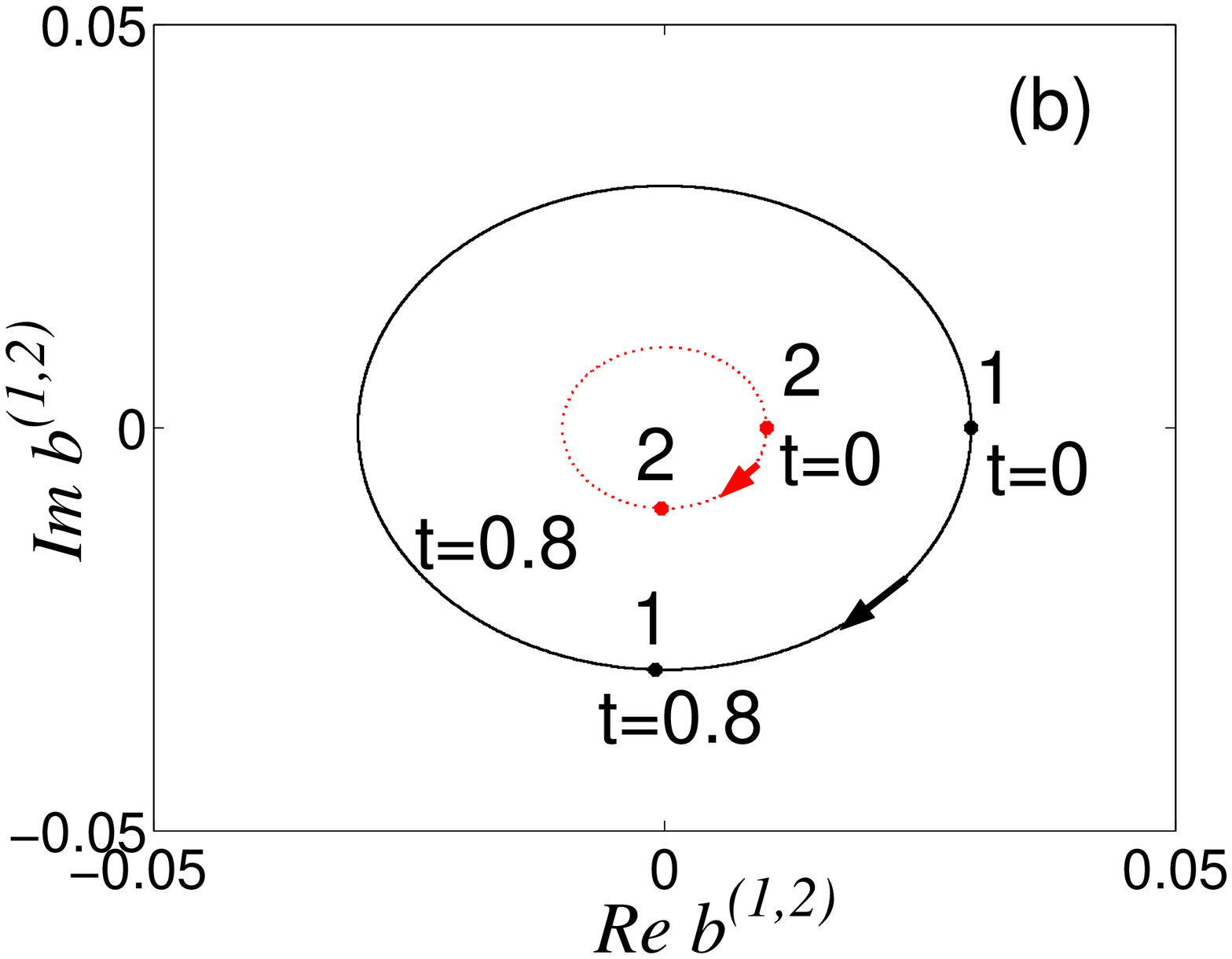}
\includegraphics[width=4cm,height=4cm,angle=0]{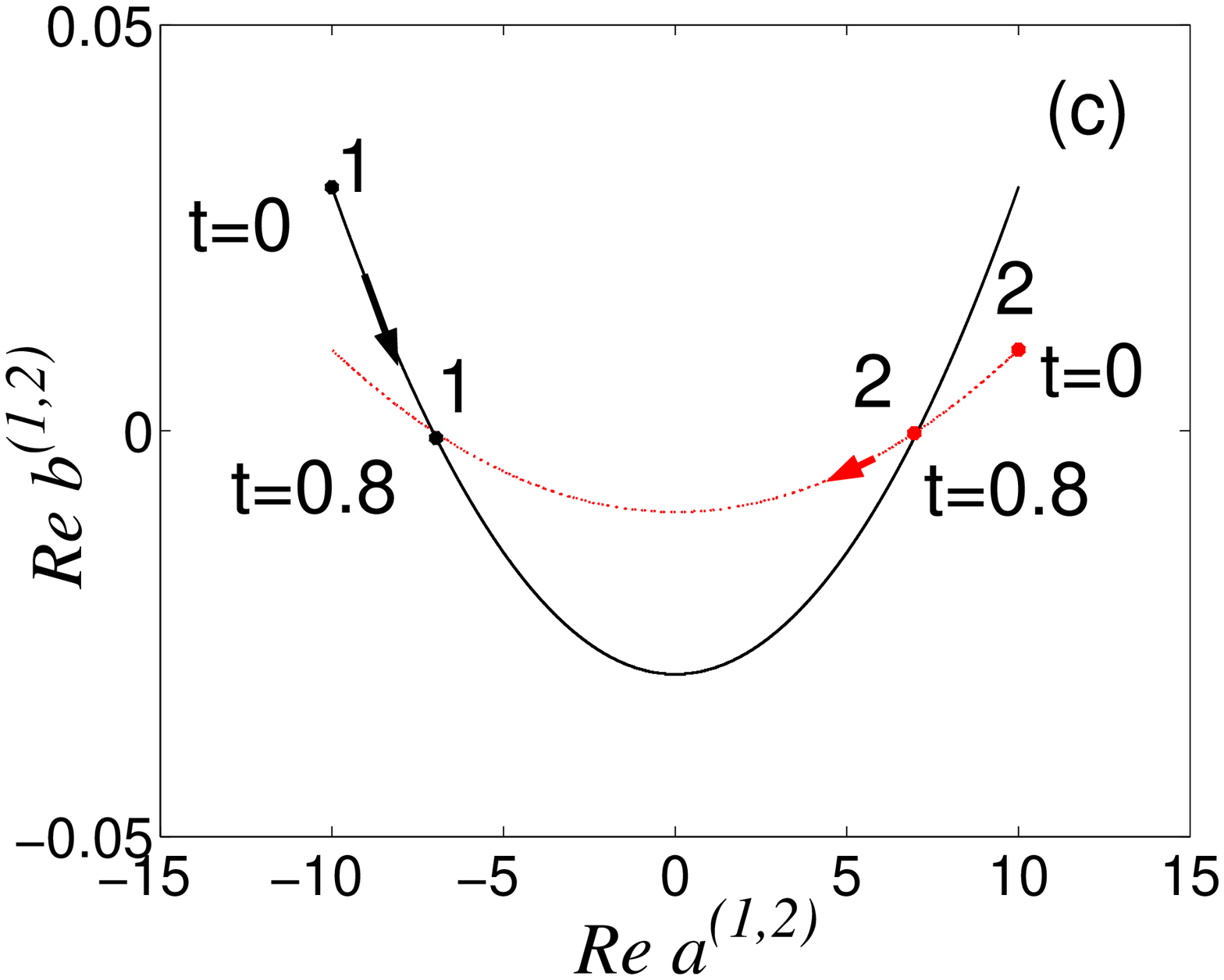}
\includegraphics[width=4cm,height=4cm,angle=0]{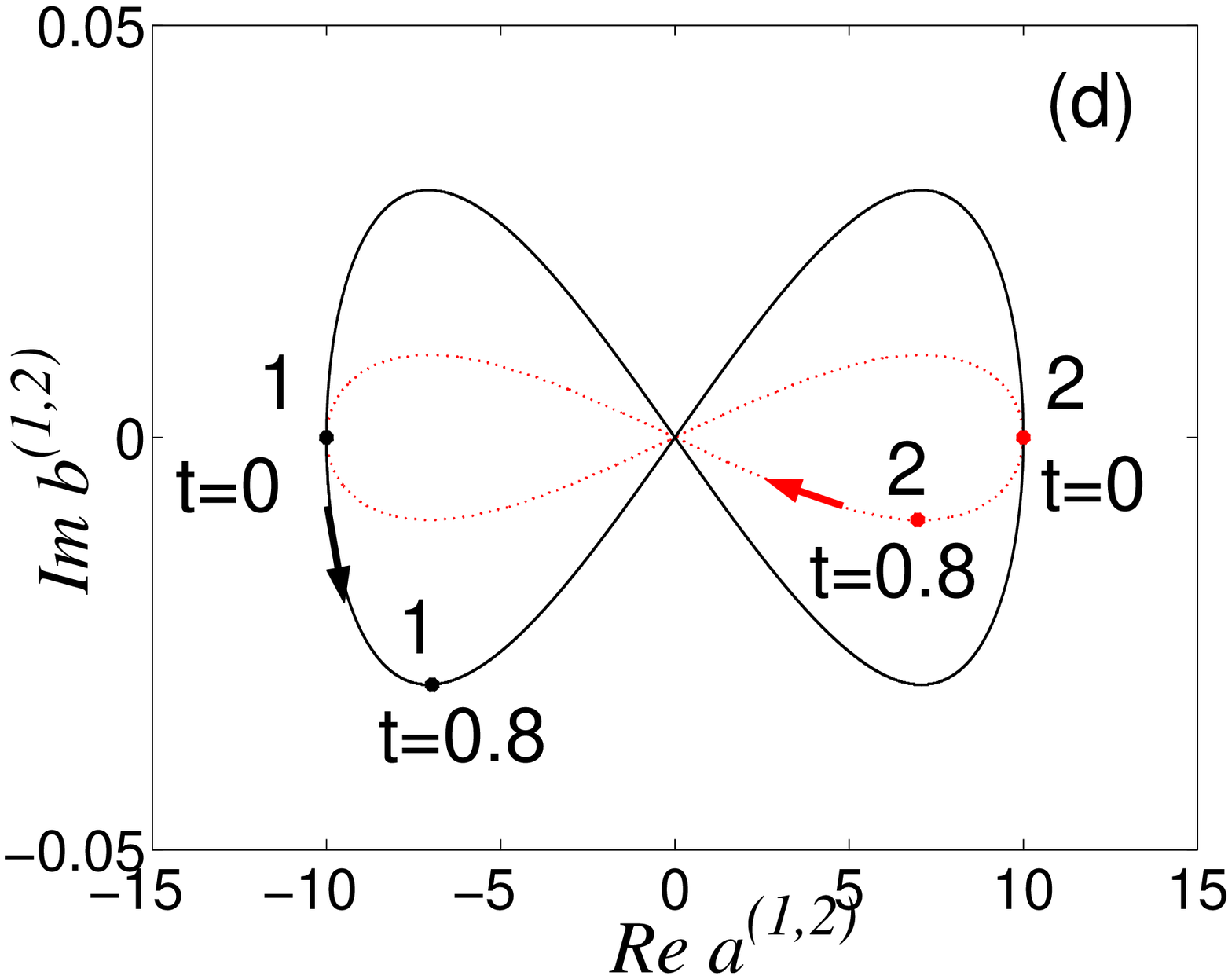}
\includegraphics[width=4cm,height=4cm,angle=0]{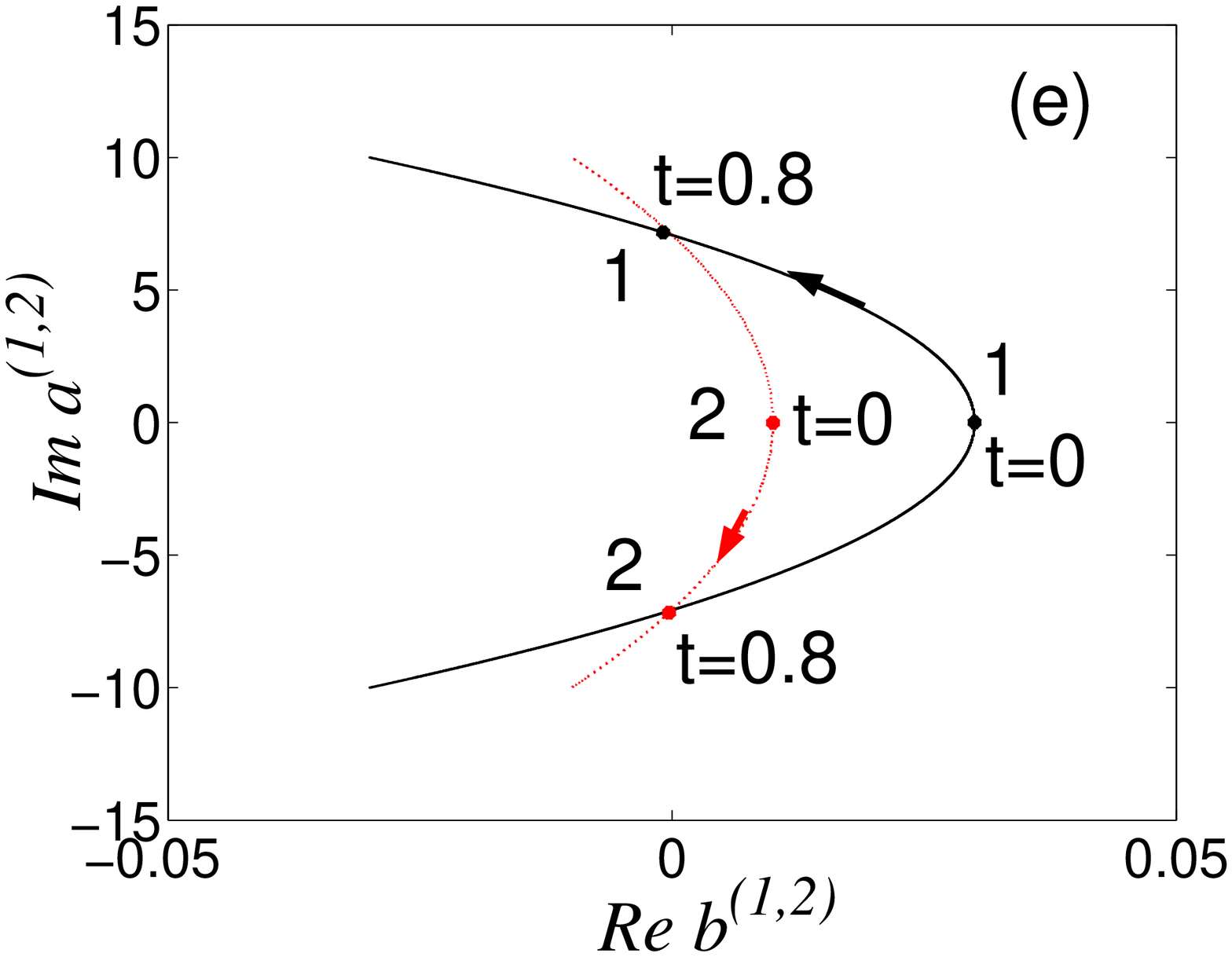}
\includegraphics[width=4cm,height=4cm,angle=0]{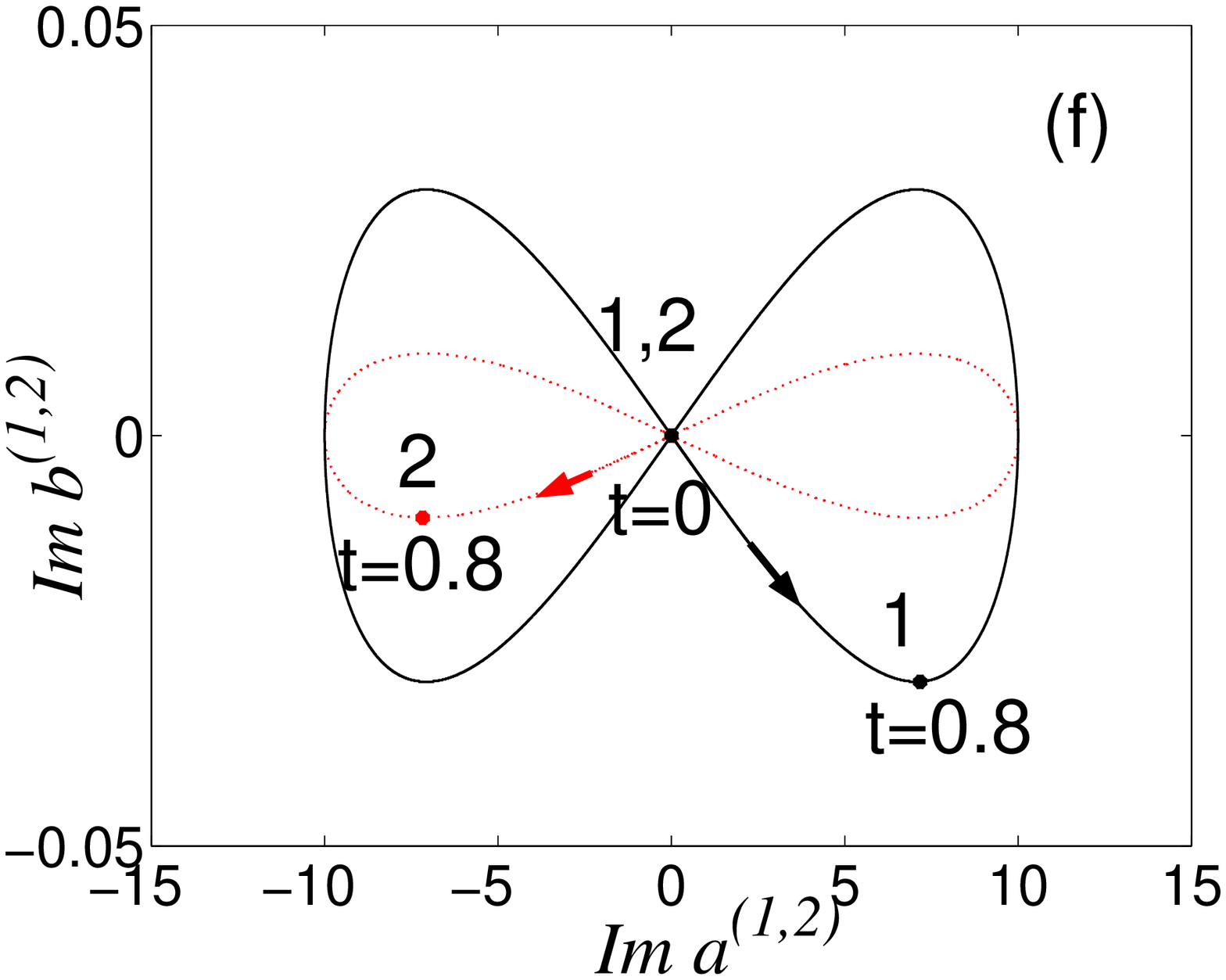}
\caption{Phase-diagrams of the coexisting solutions (\ref{non7})--(\ref{non8})
for $\omega=1$, $\epsilon=0.1$, $\gamma_{1}=0.002$, $F_{1}=0.01$ and $F_{2}=5$. }
\label{fig.3}
\end{figure}

\subsection{Fractional resonance}

It is assumed that the difference between the periodic solutions of
the autonomous and the nonautonomous system is mainly that the solution of the former
has the period (frequency) being a function of the initial conditions and the
parameter governing the nonlinearity of the system itself
(vide Eqs.(\ref{e3})--(\ref{e6})), whereas that of
the latter have period of the external pump fields only (vide Eqs.
(\ref{non7})--(\ref{non8})).% and the nonlinearity is hidden in amplitudes.
The main difference between the periodic solutions of the autonomous and nonautonomous systems
is that the period of the former is a function of initial conditions and the parameter governing
the nonlinearity of the system (vide Eqs.(\ref{e3})--(\ref{e6})), while that of the latter is
determined by the external pump fields only (vide Eqs.
(\ref{non7})--(\ref{non8})).
 However, in some cases the periodic solution of the
nonautonomous system may have the period dependent also
on the parameter governing the nonlinearity of the
differential equation. This takes place in a special case of the resonance,
 namely when  we want a nonautonomous system to vibrate at
the frequency of its autonomous counterpart.
 Then, instead of looking for the solutions of
(\ref{e1})--(\ref{e2}) in the form of (\ref{solll1})
  we search for solutions given by (see Eqs.(\ref{e3})--(\ref{e6}}))
\begin{eqnarray}
\label{sol}
a(t)=xe^{-i\Omega_{1} t}\,,\,\,\,\,
b(t)=\mp\frac{i}{2}\sqrt{\frac{x^{3}}{x^{*}}}e^{-i2\Omega_{1} t}\,,
\end{eqnarray}
where $\Omega_{1}=\omega\pm\frac{1}{2}\epsilon\sqrt{x^{*}x}$.
It is easy to note that the functions (\ref{sol}) satisfy Eqs.(\ref{e1})-- (\ref{e2})
provided that $\gamma_{2}=0$, $F_{2}=0$ and
$x=F_{1}/\gamma_{1}$. In this way the frequency $\Omega_{1}$
becomes additionally a function of the damping
constant $\gamma_{1}$ and the amplitude $F_{1}$.
 By way of example, for $\gamma_{2}=0$, $F_{2}=0$, $\omega=1$,
 $\epsilon=0.1$ and $F_{1}=5$ the system (\ref{e1})-- (\ref{e2})
 has a periodic solution provided that $\Omega_{1}=1\pm\frac{1}{2}$.
Therefore, we have  two sets equations.
The first
\begin{eqnarray}
\label{p1}
\frac{da}{dt}&=&-i a -0.5a +0.1 a^{*}b +5e^{-i\frac{3}{2}t}\,,\\
\label{p2}
\frac{db}{dt}&=&-i2b-0.5\epsilon a^{2}\,,
\end{eqnarray}
where the solutions are given by $a(t)=10\exp(-i\frac{3}{2}t)$ and $b(t)=-5i\exp(-3it)$
and, the second
\begin{eqnarray}
\label{p3}
\frac{da}{dt}&=&-i a -0.5a +0.1 a^{*}b +5e^{-i\frac{1}{2}t}\,,\\
\label{p4}
\frac{db}{dt}&=&-i2b  -0.5\epsilon a^{2}\,,
\end{eqnarray}
where $a(t)=10\exp(-i\frac{1}{2}t)$ and $b(t)=5i\exp(-it)$.
Both sets of equations
describe the so-called fractional (subharmonics, demultiplication) resonances.
%Obviously, both sets
The periodic  solutions of (\ref{p1})--(\ref{p2}) and (\ref{p3})--(\ref{p4})
satisfy  the conservative autonomous system
$da/dt=-ia+0.1a^{*}b$ and $db/dt=-i2b-0.5a^{2}$ and have phase representation
identical to that in Fig.(\ref{fig.1}).
Finally let us note that the system (\ref{e1})--(\ref{e2}) if
 $\gamma_{2}=0$, $F_{2}=0$ and the resonance condition $\Omega_{1}=\omega$ holds,
 has a Bloembergen-type solution in the form:
\begin{eqnarray}
\label{bl7} a(t)&=&\frac{F_{1}}{\gamma_{1}}\, sech \left(
\frac{\sqrt{2}}{2} \frac{F_{1}}{\gamma_{1}} \epsilon t \right)
e^{-i\omega t}\,,
\end{eqnarray}
\begin{eqnarray}
\label{bl8}
b(t)=-\frac{F_{1}}{\gamma_{1}\sqrt{2}}\tanh\left(\frac{\sqrt{2}}{2}\frac{F_{1}}{\gamma_{1}}
\epsilon t \right) e^{-i2\omega t}\,.
\end{eqnarray}
The above functions for $t\rightarrow \infty$ tend to periodic states.

\section{Chaotic behaviour}

The coexisting periodic solutions (\ref{non7})--(\ref{non8})
naturally lead to the question about the effects of a disturbance of the resonance conditions.
It is intuitively clear that this problem can only be solved numerically.
 As a background to numerical investigations we use the equations
\begin{eqnarray}
\label{e1n}
\frac{da}{dt}&=& -ia-0.002a+0.1a^{*}b +0.01e^{-i\Omega_{1}t}\,,\\
\label{e2n}
\frac{db}{dt}&=&-i2b-0.05a^{2}+5e^{-i\Omega_{2}t}\,,
\end{eqnarray}
where $\Omega_{1}$ and $\Omega_{2}$ play a role of parameters.
If at the time $t=0$ the state of the above system is determined by the
 initial conditions $a(0)=10$ and $b(0)=0.01$ and the
 conditions of resonance are satisfied i.e. $\Omega_{1}=1$ and $\Omega_{2}= 2$,
  then the pair of periodic functions
\begin{eqnarray}
\label{sol1}
a(t)=10e^{-it}\,,\,\,
\,\,\, b(t)=0.01e^{-i2t}.
\end{eqnarray}
satisfy the differential equations (\ref{e1n})--(\ref{e2n}).
 Phase diagrams of the functions (\ref{sol1})
 are given in Fig.3 (black lines). For the initial conditions
 $a(0)=-10$ and  $b(0)=0.03$ we get the second pair of coexisting periodic
 solutions (red line in Fig.3)
\begin{eqnarray}
\label{sol2}
  a(t)=-10e^{-it} \,,\,\,\,\, b(t)=0.03e^{-i2t}
\end{eqnarray}
 Let us now  consider the behaviour of the system in the neighbourhood
 of the resonance periodic solutions
 in the range
\begin{eqnarray}
\label{rez1}
\Omega_{1}=1\,\,\,\, \mbox{and}\,\,\,\, 0<\Omega_{2}<4\,.
\end{eqnarray}
 Numerically, we solve Eqs.
(\ref{e1n})--(\ref{e2n}), with the help of a fourth-order Runge-Kutta method,
 with the initial conditions $(a(0)=10$, $b(0)=0.01)$,
for $\Omega_{1}=1$ and selected values of $\Omega_{2}\neq 2$. Then, we repeat the same
procedure with the initial conditions $(a(0)=10$, $b(0)=0.03)$.
The numerical results are reflected in phase portraits
and in the spectra of Lyapunov exponents, computed by the
 Wolf procedure \cite{wolf}.

  At the beginning, let us consider two the most typical types of behaviour  being a result
of detuning, that is the resonance condition breaking.\\

  {\bf Example I - large frequency detuning}.\\
   Take the frequency  $\Omega_{2}=0.472$ instead of $\Omega_{2}=2$.
    We observe that initially the phase point
  follows the periodic solution (\ref{sol1}) and then escapes from it to
  a chaotic state. This result of the large detuning is seen in  the phase
$(\mathit{Im}\,a\,,\mathit{Re}\, a)$ plane Fig.\ref{fig.4}(a).
A symmetric behaviour is observed in the same phase portrait, if the system
(\ref{e1n})--(\ref{e2n}) starts with
the initial conditions $a(0)=-10$ and $b(0)=0.03$,
see Fig.\ref{fig.4}(b).
  The periodic ($\Omega_{2}=2$) and chaotic ($\Omega_{2}=0.472$)
  states of the system presented in Fig.\ref{fig.4}
  confirm the  spectra of the Lyapunov exponents$\{-0.0007,-0.0007,-0.0022,-0.0022\}$ and
   $\{0.1659,0.0012,-0.0029,-0.1699\}$, respectively. The latter spectrum, containing
 two positive exponents, indicates   a strong chaotic behaviour, the so-called hyperchaos.\\
%____________________________________________________________________________________
\begin{figure}
\includegraphics[width=7cm,height=7cm,angle=0]{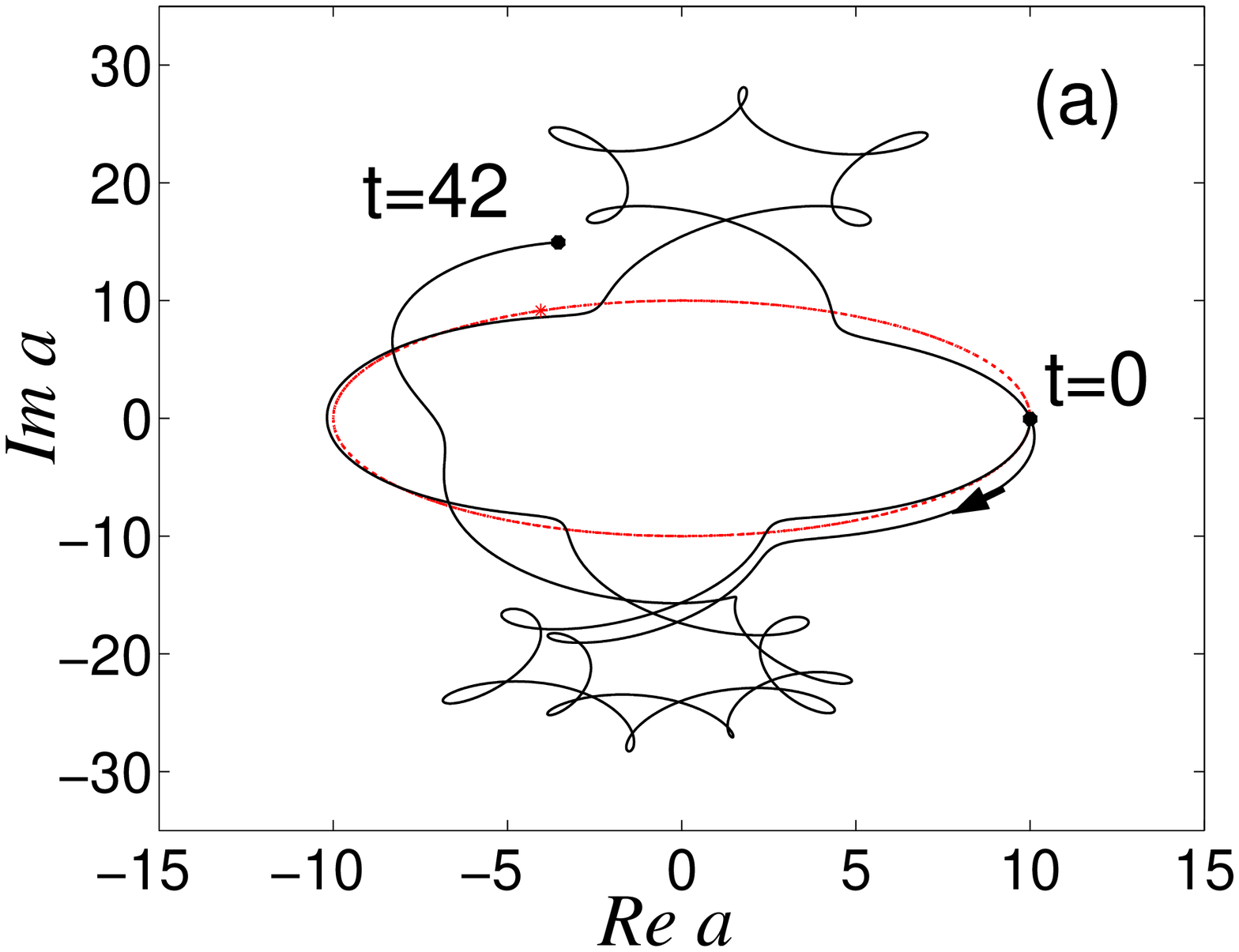}
\includegraphics[width=7cm,height=7cm,angle=0]{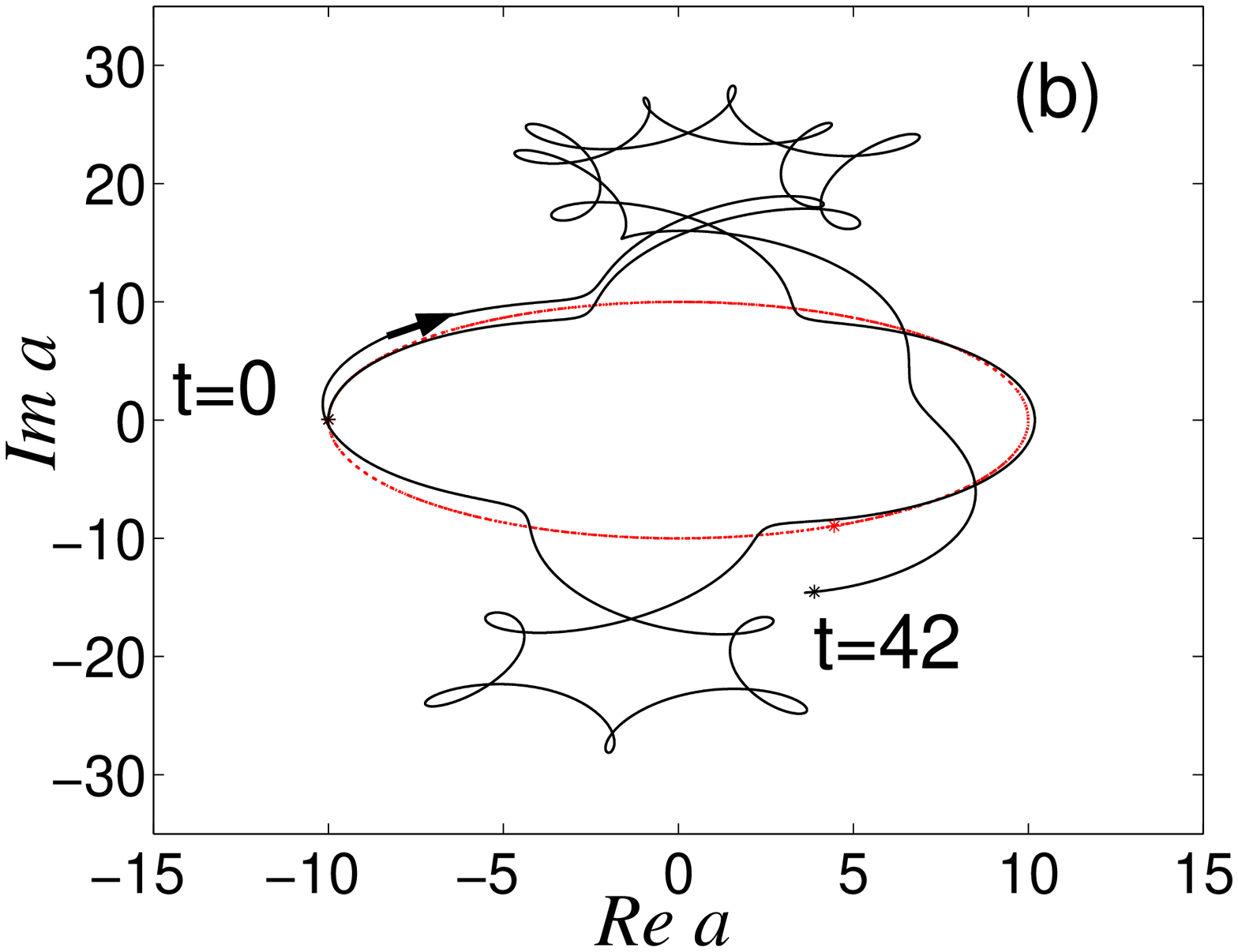}
\caption{ The symmetric escape of the phase points from the coexisting orbits.
 Figure (a) - evolution of the system (\ref{e1n})--(\ref{e2n}) with the initial
 conditions $a(0)=10$, $b(0)=0.01$ and for 1) $\Omega_{1}=1$, $\Omega_{2}=2$
 (periodic solution - read)
  and 2) $\Omega_{1}=1$,  $\Omega_{2}=0.472$ (chaos - black). Figure (b)-
  the same as in Figure(a) but with the initial conditions $a(0)=-10$ and $b(0)=0.03$.}
 \label{fig.4}
\end{figure}
%_____________________________________________________________________________________
{\bf Example II - weak frequency detuning}
\\Take now the frequency $\Omega_{2}=2.25$ instead of $\Omega_{2}=2$.
 Here, the point escapes from the periodic state
   (governed by Eqs.(\ref{sol1}))to another periodic state characterized by the  spectrum
    $\{-0.0013,-0.0014,-0.0015,-0.0015\}$. This is shown in (Fig.\ref{fig.5}).
 The same phase structure
  is obtained
   if the system starts from the coexisting periodic states
   described by Eqs.(\ref{sol2})).\\
  %_____________________________________________________________________________________
\begin{figure}
\includegraphics[width=7cm,height=7cm,angle=0]{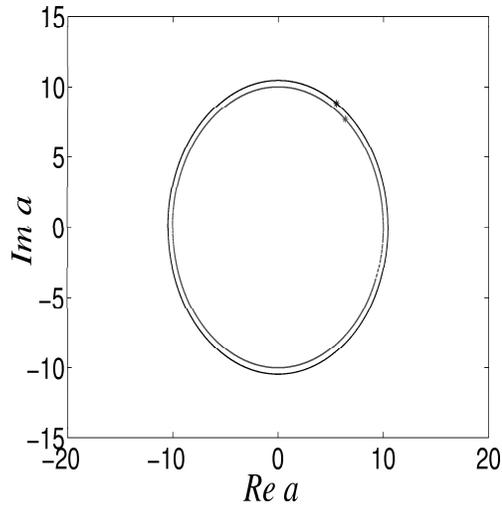}
\caption{Escape from one periodic state (red) to another
periodic state (black). Evolution of the system (\ref{e1n})--(\ref{e2n})
 with the initial conditions
$a(0)=10$ and $b(0)=0.01$ and for 1) $\Omega_{1}=1$, $\Omega_{2}=2$ (red)
and 2) $\Omega_{1}=1$, $\Omega_{2}=2.25$ (black). Black orbit has been  drawn for $t>4500$,
  to omit the transient effects).
   }
 \label{fig.5}
\end{figure}
%_______________________________________________________________________________________
Globally, the behaviour of the system in  the resonance neighbourhood
 (\ref{rez1})  is presented in
Fig.\ref{fig.6}. The  spectra of the Lyapunov exponents $\{\lambda_{1},\lambda_{2},
\lambda_{3},
\lambda_{4}\}$ versus $\Omega_{2}$ show the regions of order or chaos. If $\lambda_{1}>0$ then system is chaotic (black colour),
if simultaneously $\lambda_{1}>0$
and $\lambda_{2}>0$  then system is hyperchaotic (black and green) and finally if
$\lambda_{1}\leq 0$ the system behaves nonchaotically (periodically).
 For $\Omega_{2}=2$  the system is in a periodic state (Eqs.(\ref{sol1}) or
solid lines in Fig.\ref{fig.3}).\\
Fig. (\ref{fig.6})  shows that the spectra are nearly symmetric relative to
the resonance frequency $\Omega_{2}=2$.
 By way of example, for $\Omega_{2}=2.25$ (see Fig.\ref{fig.5}) we have the
  same spectrum as for
$\Omega_{2}=1.75$ and get two identical phase structures.

\begin{figure}
\includegraphics[width=7cm,height=7cm,angle=0]{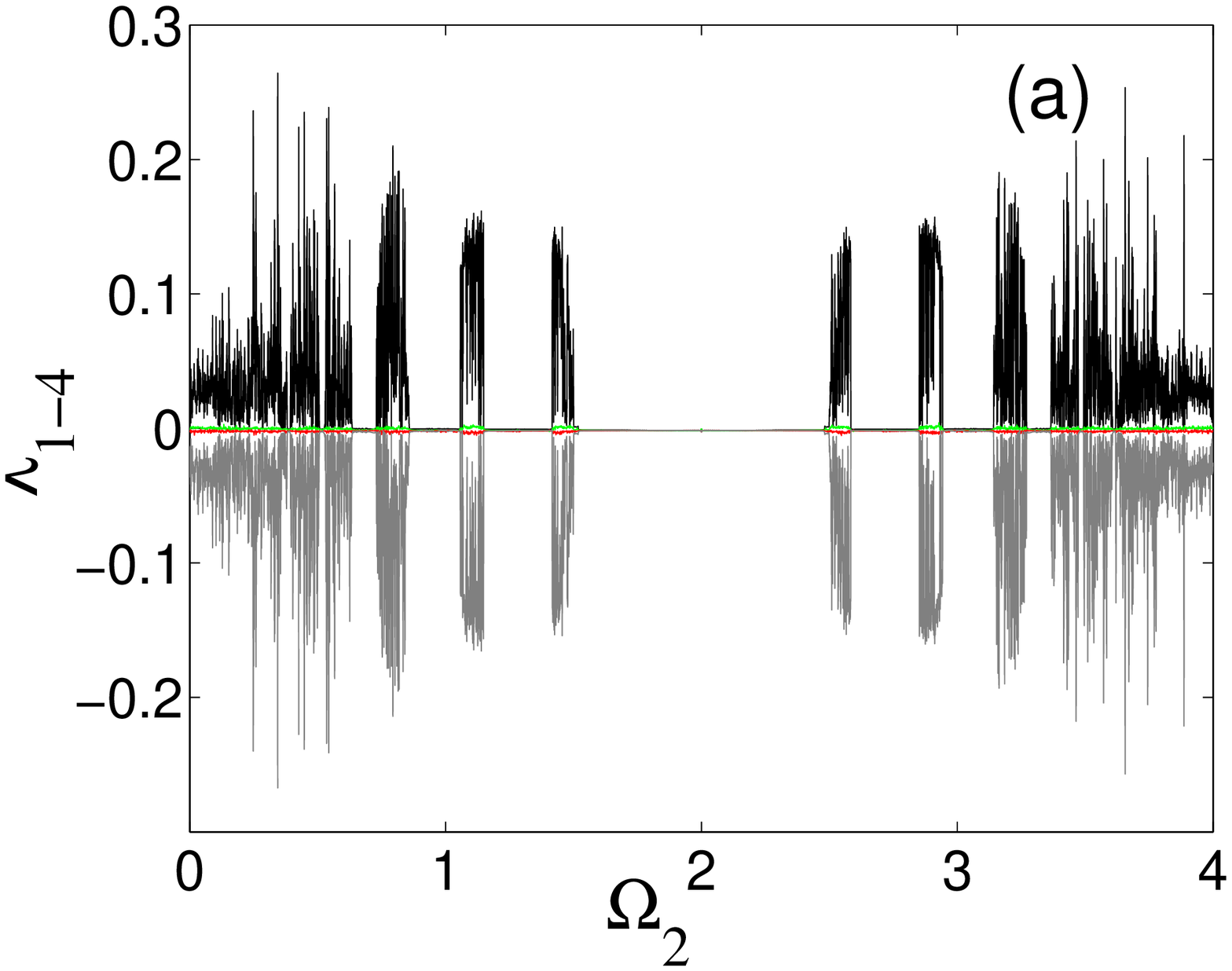}
\includegraphics[width=7cm,height=7cm,angle=0]{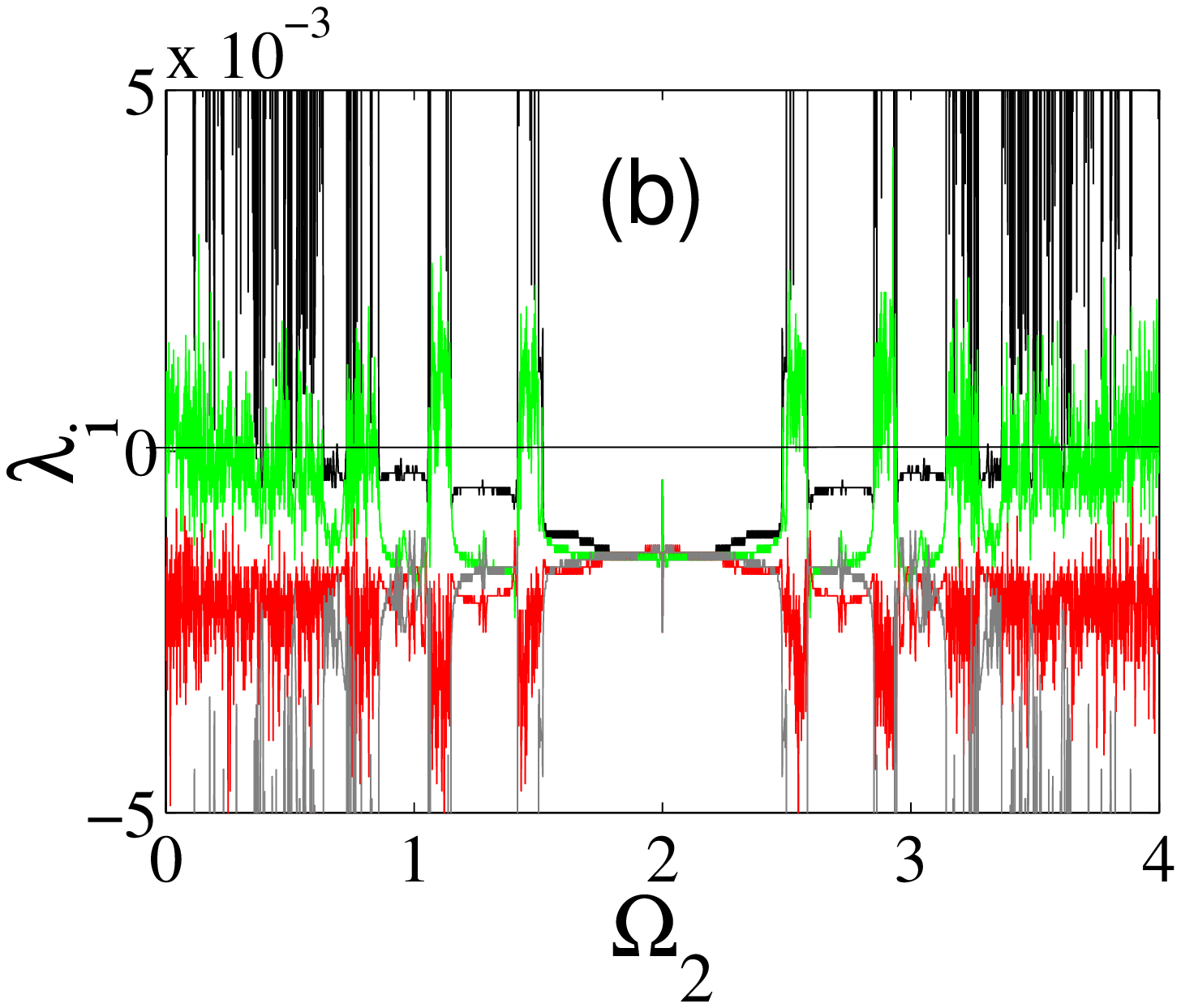}
\caption{(a) Spectrum of Lyapunov exponents $\lambda_{i}$,
$i=1,2,3,4$, for the system (\ref{e1n})--(\ref{e2n}) with the
initial conditions $a(0)=10$ and $b(0)=0.01$, for $\Omega_{1}=1$,
and  $0<\Omega_{2}<4$. (b) An enlargement of the region of $\mid
\lambda_{j}\mid < 5\times 10^{-3}$.} \label{fig.6}
\end{figure}

  A convenient way of finding out what may be expected in the system
   (\ref{e1n})--(\ref{e2n}) under conditions $\Omega_{1}\neq 1$ and
 $\Omega_{2}\neq 2$ is to calculate the maximal Lyapunow exponent
 $\lambda_{1}$ as a function of $\Omega_{1}$ and $\Omega_{2}$.
 Then, the global  dynamics of the system is simply presented
 by the Lyapunov map in the space of ($\Omega_{1}, \Omega_{2}$) (Fig.\ref{fig.7}),
 where the values  of $\lambda_{1}$ are marked by an appropriate colour.
 A band structure
 of the map shows that the system is much more sensitive to the changes in $\Omega_{2}$
 than to those in $\Omega_{1}$. This is caused by the fact that the amplitude
 of the forcing term in (\ref{e1n}) is $500$ times lower
 than in the forcing term in (\ref{e2n}). Therefore, the changes in $\Omega_{1}$
 do not affect so much the stability of the system as those in $\Omega_{2}$.
 The map is simply symmetric to the line $\Omega_{1}=1$ which means that
 the detuning $\Omega_{2}\ + \delta$  as well as $\Omega_{2}\ - \delta$ causes
 the same kind of stability (instability) of the system. The cental point of
 the map $(\Omega_{1}=1, \Omega_{2}=2)$ corresponds to the periodic solution
 (\ref{sol1}).\\
We might expect that if  the nonlinear interaction is
sufficiently weak, that is when $\epsilon/\omega <<1$, then in the presence of detuning
(that is in the neighbourhood of the  periodic solution (\ref{sol1}))
beats could appear. Though the nonlinear interaction in (\ref{e1n})--(\ref{e2n}) can really
 be treated as weak $(\omega=1, \epsilon =0.1)$ it is not sufficiently weak to induce
 quasiperiodic  phenomenon. Distinct beats  appear in (\ref{e1n})--(\ref{e2n}) when
 $\epsilon/\omega <0.01$. This condition is satisfied (with a wide margin) if we put
  in (\ref{e1n})--(\ref{e2n}),
 for example:  $\Omega_{1}=\omega=100$ (instead of $\omega=1$)
and $\Omega_{2}=200\pm \delta$, where $\delta$ is the value of detuning
 (for $\delta=0$ the system has the periodic solutions redefined $a(t)=10e^{-i100t}$ and
$b(t)=0.01e^{-i200t}$). A typical example
of beats is presented in Fig.\ref{fig.8}.
 %__________________________________________________________________________________
\begin{figure}
\includegraphics[width=8cm,height=10cm,angle=0]{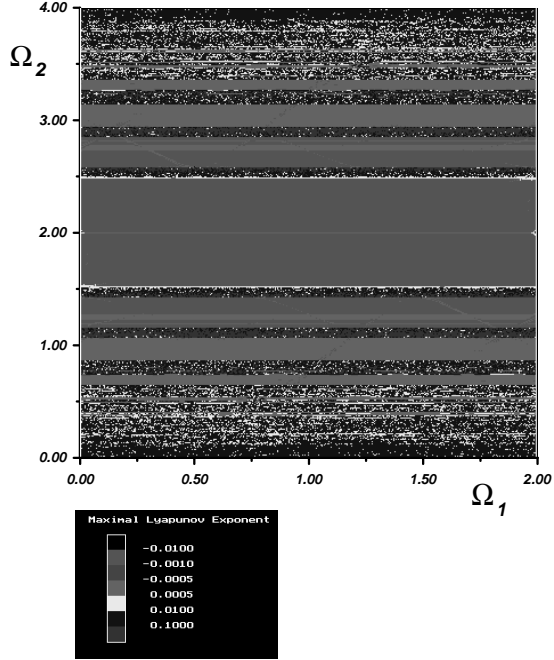}
\caption{The values of maximal Lyapunov exponents for Eqs.(\ref{e1n})--(\ref{e2n})
  with the conditions $a(0)=10$ and $b(0)=0.01$ if
 $0<\Omega<2$ and $0<\Omega_{2}<4$.}
\label{fig.7}
\end{figure}
%__________________________________________________________________________________
\begin{figure}
\includegraphics[width=5cm,height=3cm,angle=0]{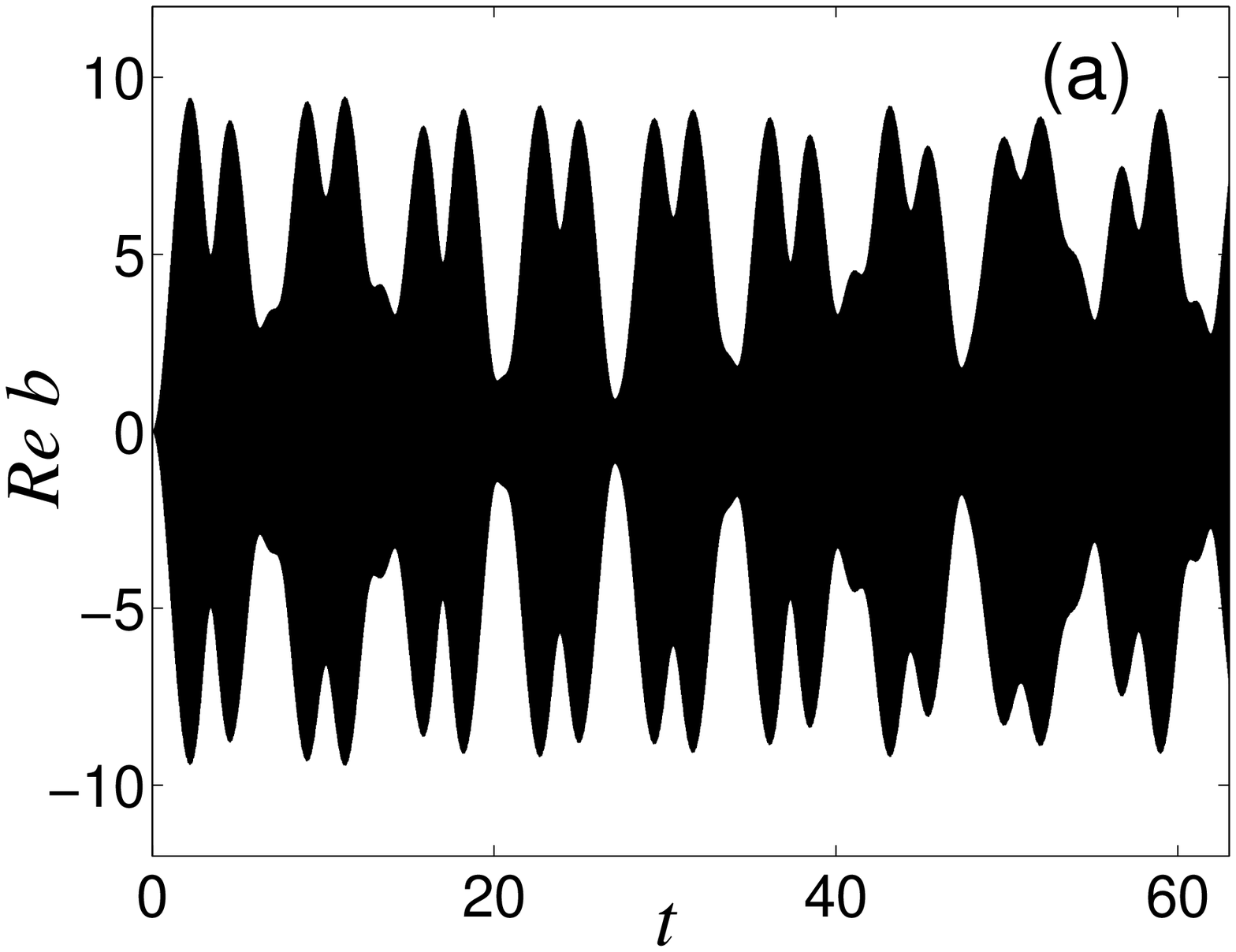}
\includegraphics[width=5cm,height=3cm,angle=0]{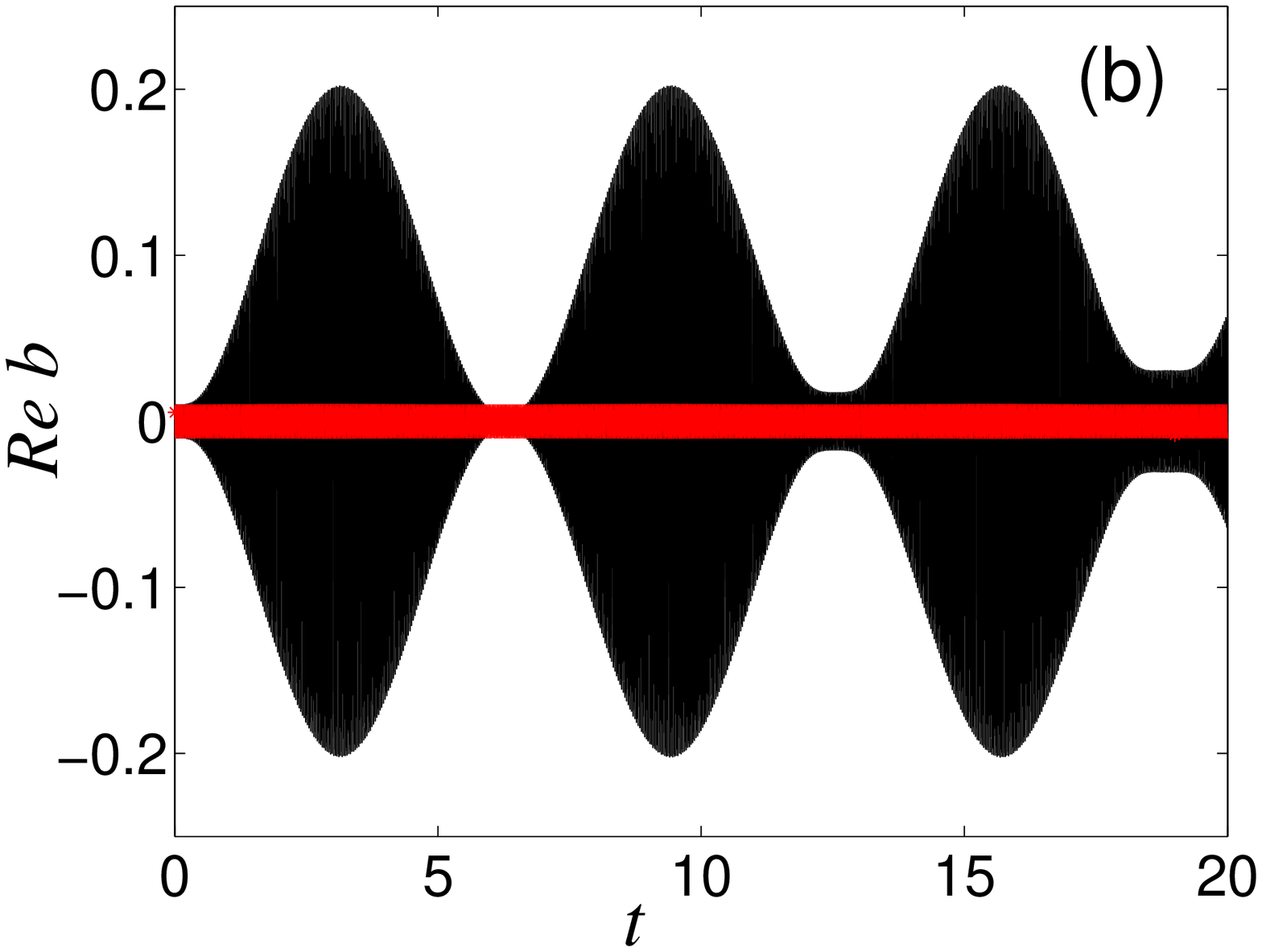}
\caption{ Beats. Time evolution $\mathit{Re}\, b(t)$ of
the system (\ref{e1})-(\ref{e2}) where $\omega=100$, $\gamma_{1}=0.002$, $\gamma_{2}=0$,
$F_{1}=0.01$, $F_{2}=5$, $\Omega_{1}=100$ and the initial conditions are
 $a(0)=10$ and $b(0)=0.01$ and for (a) $\Omega_{2}=202$, (b) $\Omega_{2}=200.1$.
The red band in (b) represents
 the periodic solution $\Omega_{2}=200$. }
\label{fig.8}
\end{figure}
 The beats proceed as follows: if $\Omega_{2}$ tends to $2\omega=200$ i.e.
($\delta \rightarrow 0$)
the amplitude of beats increases but this happens only to a certain value of the difference
$|2\omega-\Omega_{2}|=\Delta$ (in our case $\Delta=0.5$) after which the
 amplitude of beats began to
decrease. Finally  beats disappear for
$2\omega=\Omega_{2}=200$ that is in the resonance. The beats problem (also chaotic beats)
in different nonlinear systems has been recently investigated in nonlinear optics \cite{Grygiel}
and in electric circuits \cite{Cafagna}.

Nonautonomous systems do not manifest the so-called {\em translation}
 properties as their autonomous counterparts do. This property
 can be readily  reflected in phase portraits. By way of example,
 to demonstrate this behaviour we use Fig.3a (black).
 If the system (\ref{e1n})-- (\ref{e2n})
 starts from the point $A=(a(0)=10, b(0)=0.01)$ then in the phase portrait we observe
 a circle (the periodic solution (\ref{sol1})). However, if the same system starts
 from another point  $B\neq A$,
 lying on the circle, it does not remain on the circle but escapes from it and draws
 another curve  - in our  case of chaotic one (Fig.\ref{fig.9}). This chaotic behaviour
 is confirmed
 by the spectrum of the Lyapunov exponents $\{0.0059,-0.0009,-0.0031,-0.0077\}$.\\
 Generally, the system can {\em escape} from the periodic orbit to stable (periodic)
 states or to chaotic states.
 On  calculating the maximal Lyapunov exponents $\lambda_{1}$ for the system starting from individual points of
 the periodic orbit we get the information on which orbit is chaotic  $\lambda_{1}> 0$ and
 which is nonchaotic $\lambda_{1}\leq 0$. This is shown in Fig.(\ref{fig10}).
\begin{figure}
\includegraphics[width=7cm,height=7cm,angle=0]{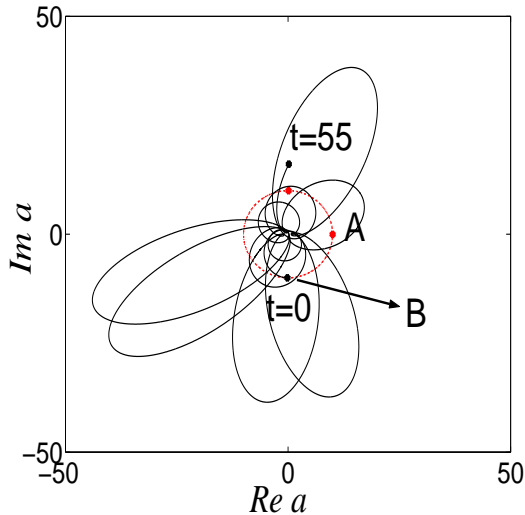}
\caption{
Escape of the phase curve starting from the  point $B$ lying
on the periodic curve (red circle).
 The periodic orbit, generated  the system (\ref{e1n})--(\ref{e2n}), starts from
 the point $A$ (the initial
conditions: $a(0)=10$ and $b(0)=0.01$).
The same system on starting from $B$ (the initial condition)
$a(0)=-0.0922-i10.0196$ and $b(0)=0.0102-i0.002$  behaves chaotically.}
 \label{fig.9}
\end{figure}

\begin{figure}
\includegraphics[width=8cm,height=8cm,angle=0]{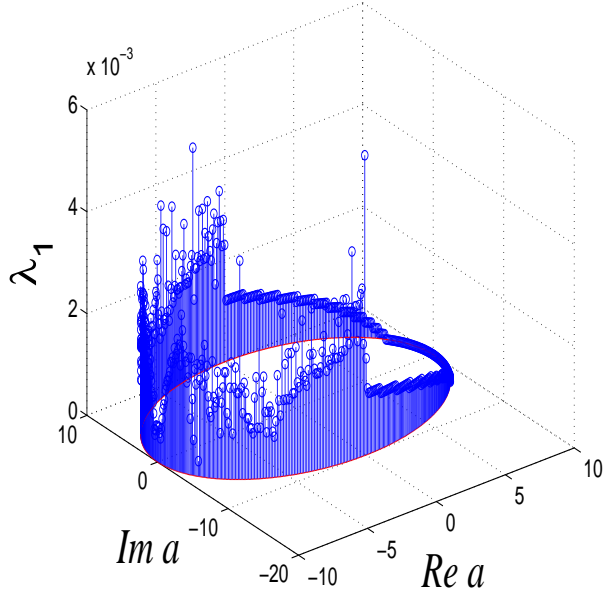}
\caption{ Maximal Lyapunov exponents $\lambda_{1}\geq 0$
for the trajectories of system (\ref{e1n})--(\ref{e2n}) started from different points
of the periodic orbit (red) in Fig.9. The individual values of $\lambda_{1}$ show
from which points (lying on the orbit) chaotic ($\lambda_{1}>0$) , and from which nonchaotic ($\lambda_{1}=0$)
phase curves originate.}
 \label{fig10}
\end{figure}

\section{Synchronization of the coexisting states. \\Quenching}

Let us now consider the synchronization problem (mutual or unidirectional)
of two dynamical systems $(a,b)$ and $(A,B)$, which we may assume to be identical
in all respects but
being in two coexisting periodic states. We take the system $(a,b)$ given by
(\ref{e1}) -- (\ref{e2}) and its copy $(A,B)$ and couple them linearly:
\begin{eqnarray}
\label{s1}
\frac{da}{dt}&=&-i\omega a -\gamma_{1}a +\epsilon a^{*}b +F_{1}e^{-i\Omega_{1}t}\\
\nonumber&-&S_{(a,A)}(a-A)
\,,\\
\label{s2}
\frac{db}{dt}&=&-i2\omega b -\gamma_{2}b -\frac{1}{2}\epsilon a^{2}+F_{2}
e^{-i\Omega_{2}t}
\\\nonumber
&-&S_{(b,B)}(b-B)\,\\
\label{ss1}
\frac{dA}{dt}&=&-i\omega A -\gamma_{1}A +\epsilon A^{*}B +F_{1}e^{-i\Omega_{1}t}\\
\nonumber &-&S_{(A,a)}(A-a)\,\\
\label{ss2}
\frac{dB}{dt}&=&-i2\omega B -\gamma_{2}B -\frac{1}{2}\epsilon A^{2}+F_{2}
e^{-i\Omega_{2}t}\,
\\\nonumber
&-&S_{(B,b)}(B-b)\,
\end{eqnarray}
where $S_{(i,j)}=S_{(j,i)}$ is a parameter of the coupling. The coupling is usually
turned on at an arbitrarily chosen time.\\
 The most spectacular behaviour is observed
if the systems $(a,b)$ and $(A,B)$ are autonomous and conservative, that is when
 $\gamma_{1}=\gamma_{2}=0$ and $F_{1}=F_{2}=0$. Suppose that the system $(a,b)$
 is in the state (\ref{e3}) -- (\ref{e4}) whereas  $(A,B)$ is in the state
 (\ref{e5}) -- (\ref{e6}). This means that the oscillator $a$ vibrates at
 the frequency $\omega+\delta$,  where
$\delta =0.5\epsilon \sqrt{\alpha^{*}\alpha}$, whereas the oscillator $A$ vibrates
at the frequency $\omega-\delta$. The oscillators $b$ and $B$ vibrate at
the frequencies $2(\omega+\delta)$, $2(\omega-\delta)$, respectively. Therefore,
the pairs$(a,b)$ and $(A,B)$ are simply  detuned  in frequencies. If the oscillator pairs
$(a,b)$ and $(A,B)$ are mutually coupled, they get synchronized. This is illustrated in
 Fig.\ref{fig.11}.
\begin{figure}
\includegraphics[width=5cm,height=4cm,angle=0]{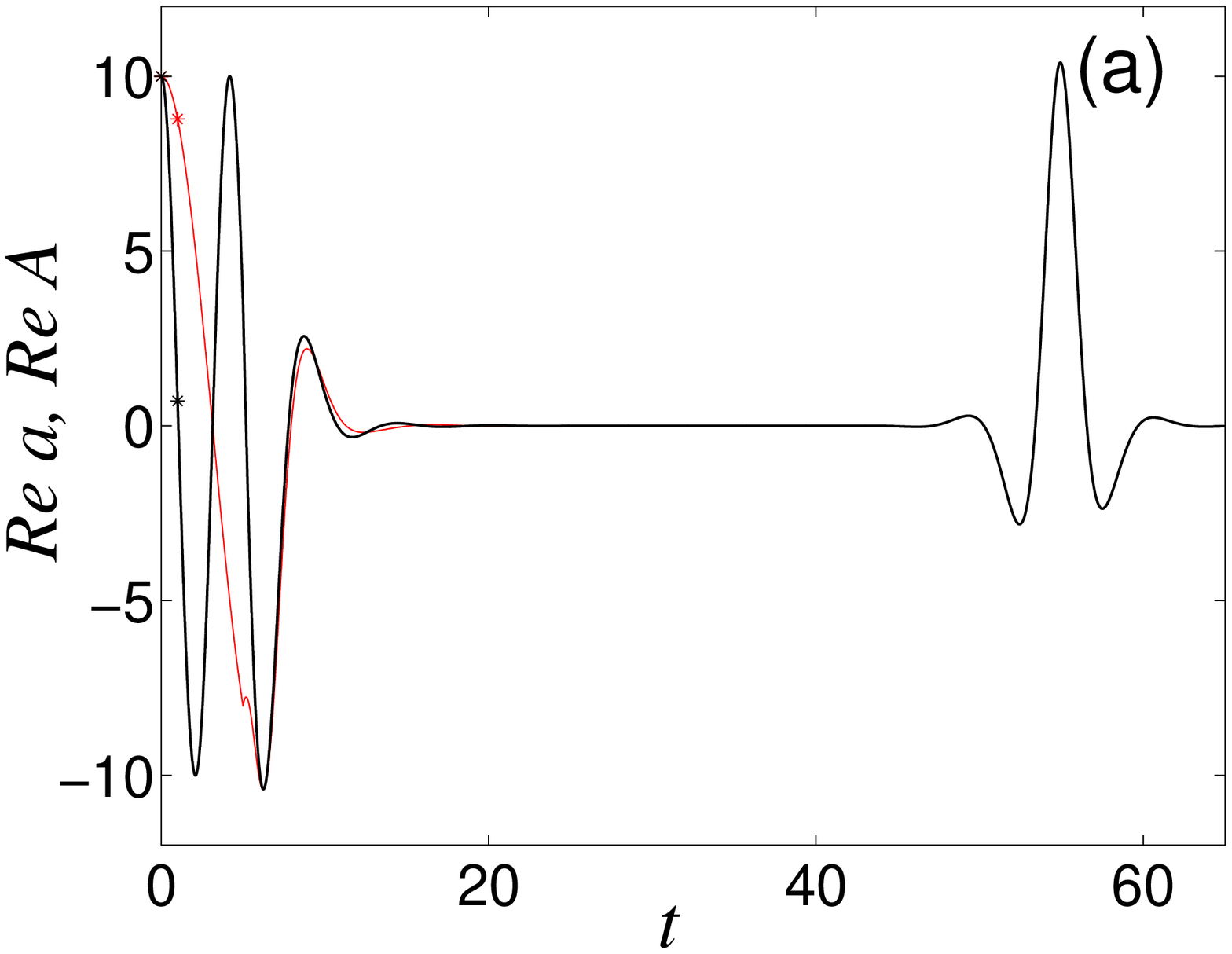}
\includegraphics[width=5cm,height=4cm,angle=0]{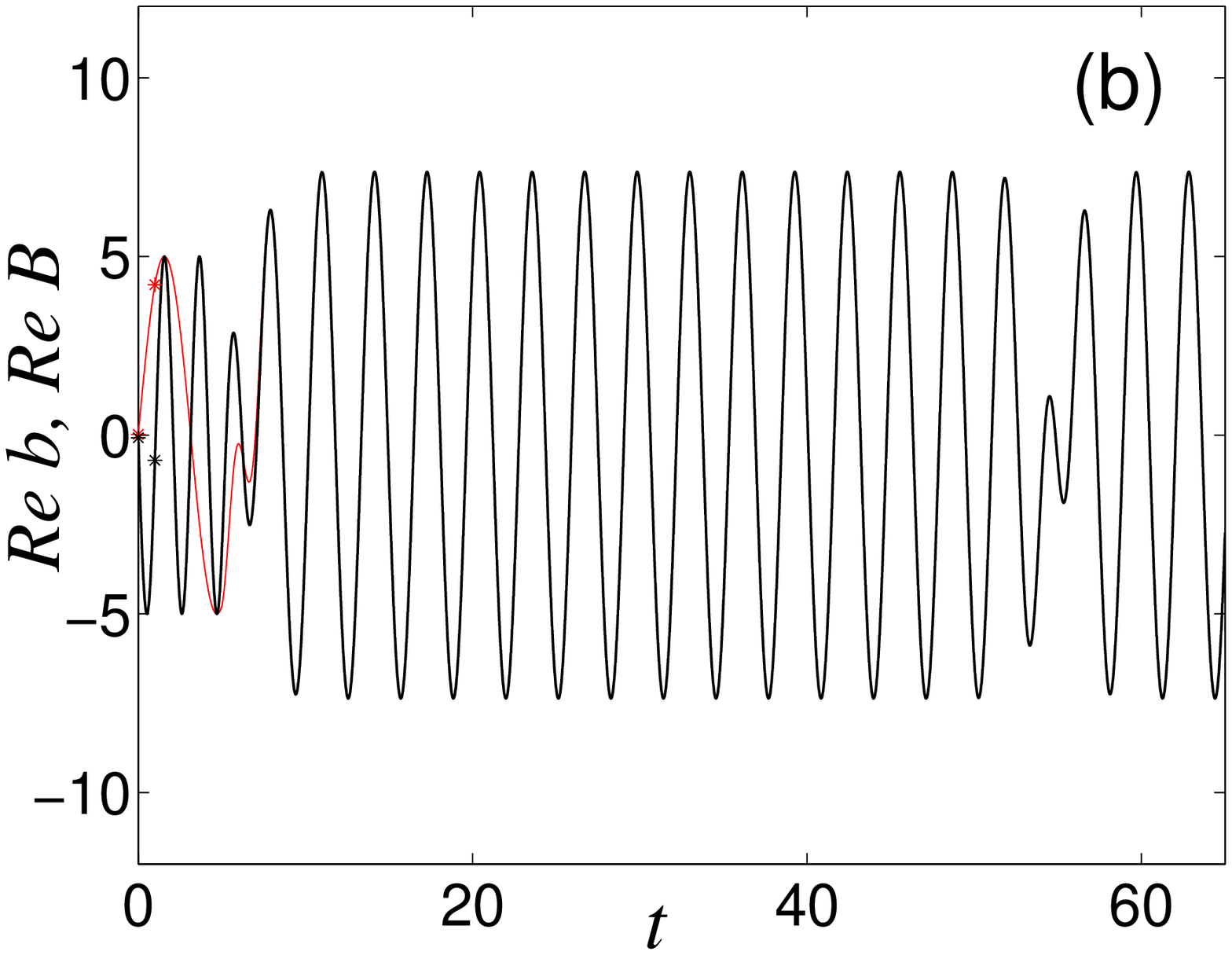}
\caption{The temporal oscillation death in $a,A$-components, black and red,
respectively (a); and oscillation in $b,B$-components (b).
The effect forced by mutual synchronization in the system (\ref{s1})--(\ref{ss2})
with the initial conditions $a(0)=10$, $b(0)=5$, $A(0)=10$, $B(0)=-5$,
for $\omega=1$, $\epsilon=0.1$, $\gamma_{1}=\gamma_{2}=0$,
 $F_{1}=F_{2}=0$ and $S_{a,A}=S_{b,B}=S_{A,a}=S_{B,b}=0.5$.
  The $S$-terms are turned at $t=10$.  }
 \label{fig.11}
\end{figure}
As seen, the coupling $S=0.5$ is turned on at $t=10$ and the systems
 synchronize at $t=15$. Consequently, for $t>15$
$a(t)=A(t)$ and $b(t)=B(t)$. What is more, the oscillator
$a$ and $A$ are quenched for some time  but at  the same time the oscillators $b$ and $B$
vibrate periodically  with the double frequency
$2\omega$. As seen from Fig.\ref{fig.11} after the oscillation death
in $a,A$-components we always observe
its rapid and short revival -- these effects  occur one after the other.
The revivals always correspond to appropriate collapses in $b,B$-components
The quenching interval depends on the coupling constant $S$, the lager the value of $S$ the
longer the quenching interval. Therefore, the quenching effect is the most
effective in the case of strong interaction that is when $S>>\omega$.
   The quenching phenomenon is forced by the
real parts of the $S_{i,j}$-terms in \ref{s1}-\ref{ss2}. They are simply  the
 momentum (velocity) terms, sometimes named diffusive \cite{Aronson},
being sources of dissipation in the individual systems.\\

Complete quenching
can be obtained in the case of  unidirectional interaction.
Suppose, that $S_{(B,b)}=S_{(A,a)}=0$, which means that  $(A,B)$ is a transmitter (master)
system and $(a,b)$ is a receiver (slave) system.
 Moreover, the transmitter sends a continuous signal \cite{Pyragas} that it is in
 the state described by (\ref{e5}) -- (\ref{e6}) because we choose the frequency $\omega=
\frac{1}{2}\epsilon \sqrt{\alpha^{*}\alpha}$. It is possible, for example,
if $\omega=0.5$, $\alpha=10$, $S_{a,A}=S_{b,B}=0.5$ and $\epsilon=0.1$ (Fig.\ref{fig.12}).
 Physically, it means that the transmitter does not vibrate. The
receiver is in a periodic state (\ref{e3}) -- (\ref{e4}). On turning on the coupling
the vibrations in the receiver are quenched -- oscillation death is complete.
The total quenching can be spectacularly observed with the help of the
 phase plots in Fig.\ref{fig.1} (a). Namely, the phase point $2$ does not move, whereas
 point $1$ follows the periodic orbit. At the moment we turn on the synchronization
  mechanism and point $1$ moves slower and slower towards point $2$,
 and finally approaches the point $2$ -- the orbit is quenched. A similar behaviour
 is observed in Fig.\ref{fig.1} (b), (c) and (e). In Fig.\ref{fig.1} (d) and (f) point
 $2$ does not lie on the orbit of point $1$. Point $1$ moves up and down
 along the {\em parabole}. If we turn on the synchronization mechanism point $1$
 escapes from its orbit  and tends to point $2$ -- the orbit is quenched. Also,
 the inverse process  is possible, that is creation of an orbit for point $2$.

With the help of the unidirectional synchronization we can force the phase
point in Fig.\ref{fig.9} to return and escape chaotically from the periodic orbit.
 To do that, it is necessary to use the system(\ref{s1})--(\ref{s2})with
 $S_{(A,a)}=S_{(B,b)}=0$,
 $\gamma_{1}=0.002$ $, \gamma_{2}=0$, $F_{1}=0.01$,
$ F_{2}=5$, $\Omega_{1}=1$, $\Omega_{2}=2$ and
the initial conditions $a(0)=-0.0922-i10.0196$, $b(0)=0.0102-i0.002$ (chaotic orbit)
and $a(0)=10$ and $b(0)=0.01$ (red circle). Moreover, the  terms $S_{(a,A)}=S_{(b,B)}=0.5 $
 should be turned on at $t=10$, then we observe the situation presented
 in Fig.\ref{fig.13}.\\
Synchronization of the nonautonomous coexisting states is as easy as
that of the autonomous ones. The resultant vibrations usually appear as intricate
revivals and collapses, frequently chaotic.

\begin{figure}
\includegraphics[width=4cm,height=4cm,angle=0]{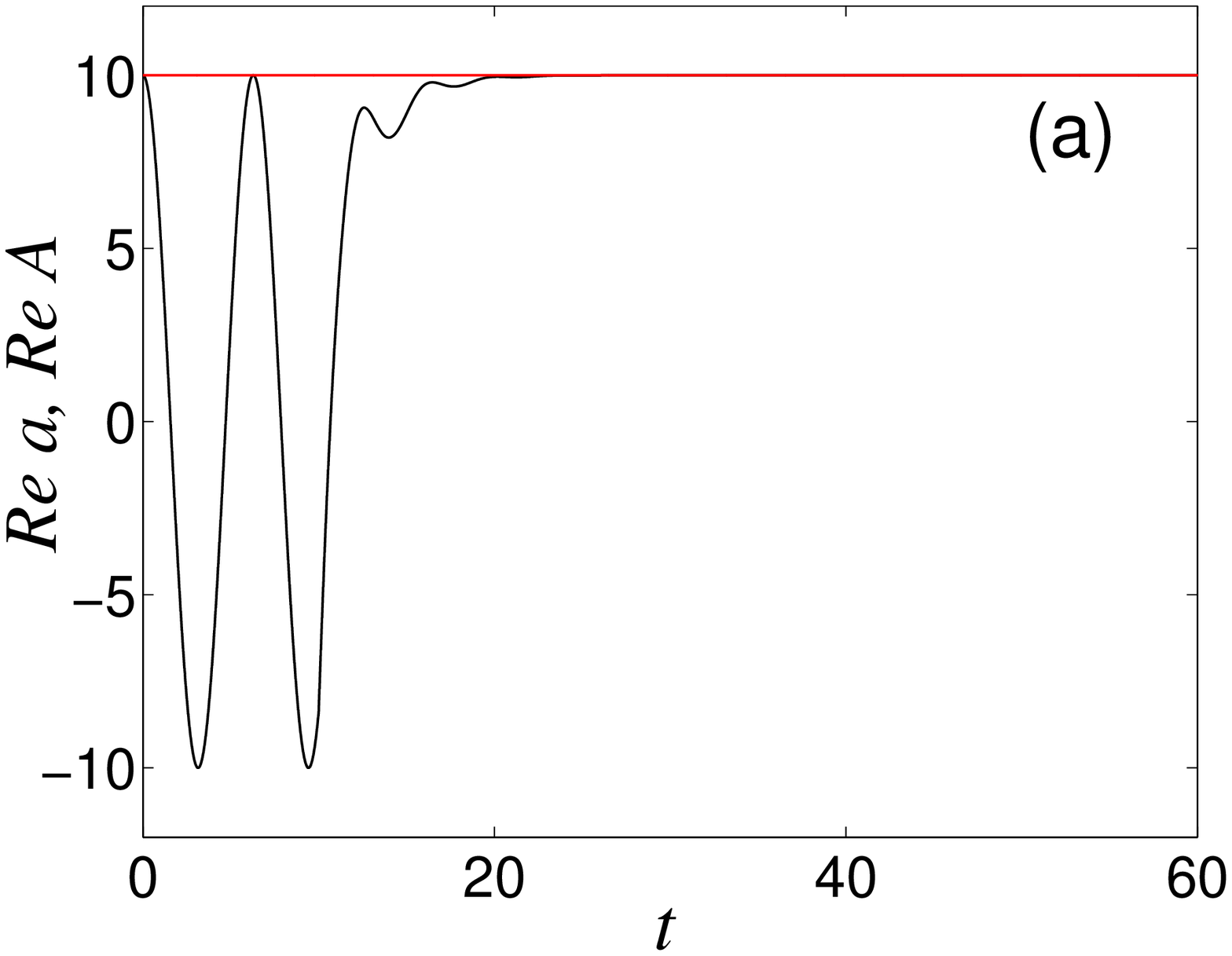}
\includegraphics[width=4cm,height=4cm,angle=0]{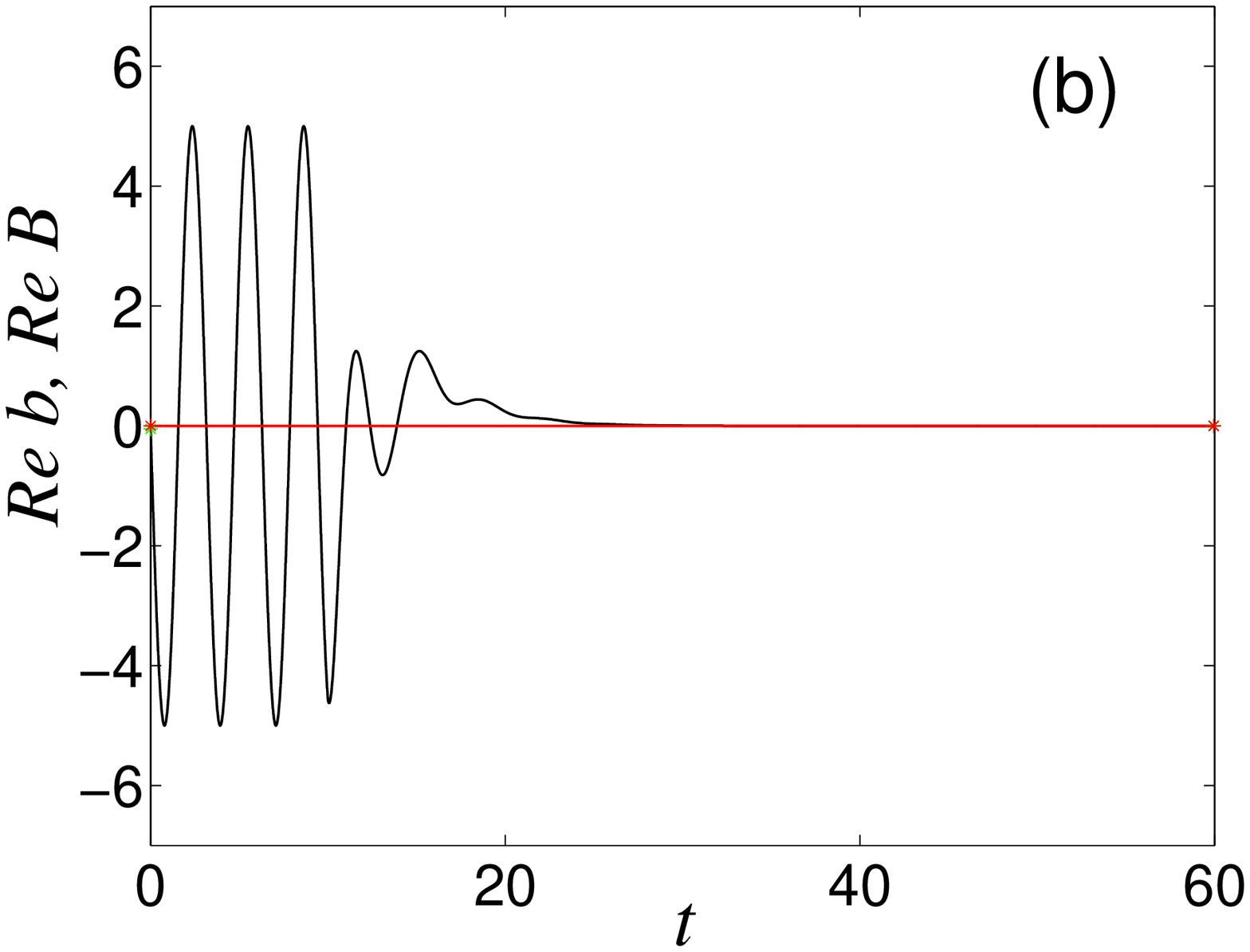}
\caption{The complete quenching caused by unidirectional  synchronization
in the system(\ref{s1})--(\ref{ss2})
with the initial conditions $a(0)=10$, $b(0)=5$, $A(0)=10$, $B(0)=-5$,
for $\omega=0.5$, $\epsilon=0.1$, $\gamma_{1}=\gamma_{2}=0$,
 $F_{1}=F_{2}=0$ and $S_{(A,a)}=S_{(B,b)}=0$. The  terms $S_{(a,A)}=S_{(b,B)}=0.5 $
 are turned on at $t=10$.}
 \label{fig.12}
\end{figure}

\begin{figure}
\includegraphics[width=4cm,height=4cm,angle=0]{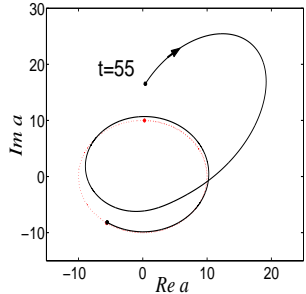}
\caption{Return to the periodic orbit of a point chaotically escaping from it. The escape
is shown in Fig.\ref{fig.9}.
The return is caused by  unidirectional synchronization switched at the time $t=55$.}
 \label{fig.13}
\end{figure}

\section{Conclusions}
The coexisting periodic solutions, presented and analyzed in this paper, are a natural
feature of the nonlinear differential equations (\ref{e1})--(\ref{e2}) describing
nonlinear optical processes of the second order.\\
The coexisting solutions of the autonomous type (\ref{e3})--(\ref{e6})
have a simple physical interpretation.  Namely, we are able  to prepare the initial states
of two independent harmonic oscillators (of the frequencies $\omega$ and $2\omega$)
in such a way that after turning on the nonlinear interaction the oscillators
vibrate at
the detuned frequencies  $\omega+\delta$ and $2(\omega+\delta)$  or at the coexisting
frequencies $\omega-\delta$ and $2(\omega-\delta)$.
  If the system is damped, this correction also
depends on the damping constant.
This is clearly seen from Eqs.(\ref{e9})--(\ref{e10}).
The coexistence in frequency but not in amplitudes seems to be characteristic
of a large class of nonlinear autonomous systems
 in the so-called rotating wave approximation (see Appendix).
Switching on a linear coupling between the two coexisting detuned systems leads to the temporal
 disappearance of $(\omega \pm \delta)$-vibrations and the appearance of
 $2\omega$-vibrations only.
The quenching interval can be controlled by the coupling parameter. In special cases
 (unidirectional coupling) we are able to completely  annihilate vibrations in both
 oscillators
 being in the coexisting states.

The physical interpretation of the coexisting solutions for the nonautonomous case
is different to that presented above. The nonlinearity is now concealed in amplitude,
and the frequency is a parameter of the system. Two periodic solutions, having
the same resonance frequency,  have different amplitudes
(or the same amplitudes but different signs). Therefore, the coexistence means
here a possibility of existence of two resonance states at the same values of parameters.
The structure of the coexisting states in the phase space is completely degenerate as in the
case of Eqs. (\ref{non1})--(\ref{non2}). The difference is only in phase,
 or partially degenerate as in the case
of solutions (\ref{non7})--(\ref{non8}). In the latter case, it means that
 in some subspace (Fig.\ref{fig.3}(a)) two
different coexisting states have identical phase curves.
Changing the external parameters of the system, for example the external pump frequency,
we can control the phase structure of the coexisting states.
 Consequently we can make the system jump from a coexisting state to a periodic,
quasiperiodic (beats) or chaotic state.

\section{Appendix}
The method presented is  a version of the Lindstedt's method \cite{minorsky}(p.224), applied to
autonomous systems in complex variables. If this method is applied to differential
eqations of nonlinear optics written in the rotating wave approximation it gives
the so-called {\em closed} solution. Otherwise, we get series solutions. Below we
 show how this method works for the system(\ref{e1a})--(\ref{e2a}). Moreover,
 we present other periodic solutions of the selected equations of nonlinear optics.

\subsection{Second-harmonic problem}
We find a periodic solution of Eqs.(\ref{e1a})--(\ref{e2a}).
The problem is considered for  arbitrary initial conditions: $a(0)=\alpha$ and
$b(0)=\beta$. It is obvious that if $\epsilon=0$ the system(\ref{e1a})--(\ref{e2a}) has
a periodic solution given  by the functions : $a(t)=\alpha \exp{(-i\omega t)}$ and
$b(t)=\beta\exp{(-i2\omega t)}$. Now, we suppose that if $\epsilon\neq 0$ the set
of equations (\ref{e1a})--(\ref{e2a}) also has periodic solution with unknown frequencies $\Omega$ and
$2\Omega$. In order to avoid dealing with the unknown frequencies in the system
(\ref{e1a})--(\ref{e2a}) we put $\Omega t=\omega \tau\,$. This leads to
\begin{eqnarray}
\label{a3}
 a(t)=a\left(\frac{\omega\tau}{\Omega}\right)=A(\tau)\,,\,\,\,\,\,
 b(t)=b\left(\frac{\omega \tau}{\Omega}\right)=B(\tau)\,,
\end{eqnarray}
and
\begin{eqnarray}
\label{a4}
\frac{da}{dt}=\frac{dA}{d\tau}\frac{\Omega}{\omega}\,,\,\,\,\,\,
\frac{db}{dt}=\frac{dB}{d\tau}\frac{\Omega}{\omega}\,.
\end{eqnarray}
On inserting (\ref{a3}) and (\ref{a4}) into (\ref{e1a})--(\ref{e2a})
we obtain the equations of motion in the new variables:
\begin{eqnarray}
\label{a5}
\frac{\Omega}{\omega}\frac{dA}{d\tau}&=&-i\omega A +\epsilon A^{*}B\,,
\nonumber\\
\frac{\Omega}{\omega} \frac{dB}{d\tau}&=&-i2\omega B -0.5\epsilon A^{2}\,.
\end{eqnarray}
In this case we can look for series solutions of the form:
\begin{eqnarray}
\label{a6}
 A&=&A_{0}+\epsilon A_{1}+\epsilon^{2}A_{2}+...\\
\label{a7}
B&=&B_{0}+\epsilon B_{1}+\epsilon^{2}B_{2}+...\\
\label{a8}
\Omega&=&\omega+\epsilon \omega_{1}+\epsilon^{2}\omega_{2}+...
\end{eqnarray}
 On substituting  (\ref{a6}),  (\ref{a7}) and (\ref{a8}) into(\ref{a5})
we get a recursive set of equations:
\begin{eqnarray}
\label{a9}
\dot{A}_{0}&=&-i\omega A_{0}\,,\\
\label{a10}
\dot{B}_{0}&=&-i2\omega B_{0\,,}\\
\label{a11}
\dot{A}_{1}&=&-i\omega A_{1} -\frac{\omega_{1}}{\omega} \dot{A}_{0}
+A^{*}_{0}B_{0}\,, \\
\label{a12}
\dot{B}_{1}& =&-2i\omega B_{1} -\frac{\omega_{1}}{\omega} \dot{B}_{0}
-0.5A^{2}_{0}\,,\\
\label{a13}
\dot{A}_{2}&=&-i\omega A_{2} -\frac{\omega_{1}}{\omega}\dot{A}_{1}
 -\frac{\omega_{2}}{\omega}\dot{A}_{0} +A_{1}^{*}B_{0} +A_{0}^{*}B_{1}\,, \\
\label{a14}
\dot{B}_{2}&=&-i2\omega B_{2}
-\frac{\omega_{1}}{\omega}\dot{B}_{1}-\frac{\omega_{2}}{\omega}\dot{B}_{0}-A_{0}A_{1}\,,\\
& &..........................................................
\nonumber
\end{eqnarray}
with the initial conditions:
\begin{eqnarray}
\label{a15}
A_{0}(0)=\alpha\,,\,\,\,\,\,\,\, B_{0}(0)=\beta\,,\nonumber\\
A_{i}(0)=0\,,\,\,\,\,\,\,\, B_{i}(0)=0\,\,\,\,\,\mbox{for}\,\,\, i>0\,.
\end{eqnarray}
The dot denotes that differentiations are with respect to $\tau$.
The zero-order solutions are:
\begin{eqnarray}
\label{a16}
A_{0}(\tau) =\alpha e^{-i\omega \tau}\,,\\
\label{a17}
B_{0}(\tau) =\beta e^{-i2\omega \tau}\,.
\end{eqnarray}
On substituting (\ref{a16}) and (\ref{a17}) into (\ref{a11}) and
(\ref{a12}) we have:
\begin{eqnarray}
\label{a19}
\dot{A}_{1}&=&-i\omega A_{1} +i\omega_{1} \alpha e^{-i\omega \tau}
+\alpha^{*}\beta e^{-i\omega \tau}\,, \\
\label{a20}
\dot{B}_{1}& =&-i2\omega B_{1} +i2\omega_{1}\beta e^{-2i\omega \tau}
-0.5\alpha ^{2} e^{-i2\omega \tau }\,.
\end{eqnarray}
The secular terms are: $i\omega_{1} \alpha e^{-i\omega \tau}
+\alpha^{*}\beta e^{-i\omega \tau}$ and $
i2\omega_{1}\beta e^{-i2\omega \tau}
-0.5\alpha ^{2} e^{-2i\omega \tau} $. To eliminate the secular terms we put
\begin{eqnarray}
\label{a20a}
i\omega_{1} \alpha+\alpha^{*}\beta=0\,,\\
\label{a20b}
 i2\omega_{1}\beta-0.5\alpha^{2}=0\,.
\end{eqnarray}
These assumptions reduce
the set of equations (\ref{a19}) -- (\ref{a20}) to the form
\begin{eqnarray}
\label{a21}
\dot{A}_{1}&=&-i\omega A_{1}\,,  \\
\label{a22}
\dot{B}_{1}&=&-i2\omega B_{1}\,.
\end{eqnarray}
The above equations with zero initial conditions
  $A_{1}(0)=0$ and $B_{1}(0)=0$ have trivial
solutions, therefore
: $A_{1}(\tau)=0$ and $B_{1}(\tau)=0$.
Now,  we calculate the new frequency
$\Omega= \omega+\epsilon \omega_{1}$. From (\ref{a20a}) and
  (\ref{a20b}) we obtain  $\omega_{1}=i
\frac{\alpha^{*}\beta}{\alpha}$ and $\omega_{1}=-\frac{i}{4}
\frac{\alpha^{2}}{\beta}$. This is only possible if $\beta=\pm
\frac{i}{2} \sqrt{ \frac{\alpha^{3}}{\alpha^{*}} }$.
 Therefore, we have $\Omega= \omega\pm 0.5\epsilon\sqrt{\alpha^{*}\alpha}$. Finally, from (\ref{a6})--(\ref{a7}) and
 (\ref{a3}) we get
\begin{eqnarray}
\label{a23}
a(t)&=&\alpha e^{-i(\omega\pm\frac{1}{2}\epsilon\sqrt{\alpha^{*}\alpha})t}
+\epsilon*0\\
\label{a24}
b(t)&=&\mp \frac{i}{2}\sqrt{\frac{\alpha^{3}}{\alpha^{*}}}e^{-i2(\omega\pm\frac{1}{2}\epsilon
\sqrt{\alpha^{*}\alpha
})t} +\epsilon *0
\end{eqnarray}
Now, we can  consider second-order corrections. Because
$A_{1}=0$ and $B_{1}=0$,  Eqs (\ref{a13})--(\ref{a14}) have
the form:
\begin{eqnarray}
\label{a25}
\dot{A}_{2}&=&-i\omega A_{2}
 -\frac{\omega_{2}}{\omega}\dot{A}_{0}= -i\omega
 A_{2}+i\omega_{2}\alpha e^{-i\omega \tau}\,, \\
\label{a26}
\dot{B}_{2}&=&-i2\omega B_{2}
-\frac{\omega_{2}}{\omega}\dot{B}_{0}=-i2\omega
B_{2}+i2\omega_{2}e^{-i2\omega \tau}\,.
\end{eqnarray}
The secular terms are: $(\omega_{2}/\omega)\dot{A}_{0}$ and
$(\omega_{2}/\omega)\dot{B}_{0}$. Therefore, we have to put
$\omega_{2}=0$ which leads to the following equations:
\begin{eqnarray}
\label{a27}
\dot{A}_{2}&=&-i\omega A_{2}\,, \\
\label{a28}
\dot{B}_{2}&=&-i2\omega B_{2}
\,.
\end{eqnarray}
The above equations also have (with zero-initial conditions) zero-solutions:
$A_{2}(\tau)=0$ and $B_{2}(\tau)=0$.
 Consequently:
\begin{eqnarray}
\label{a29}
a(t)&=&\alpha e^{-i(\omega\pm \frac{1}{2}\epsilon\sqrt{\alpha^{*}\alpha})t}
+\epsilon*0+ \epsilon^{2}*0\\
\label{a30}
b(t)&=&\mp \frac{i}{2}\sqrt{\frac{\alpha^{3}}{\alpha^{*}}}e^{-i2(\omega\pm\frac{1}{2}\epsilon
\sqrt{\alpha^{*}\alpha})t} +\epsilon *0+ \epsilon^{2} *0
\end{eqnarray}
Generally, the mathematical induction method leads to $\omega_{i}=0$
and $A_{i}(\tau)=B_{i}(\tau)=0$ for $i>1$. Therefore, the above solutions are closed.
Finally, we have:
\begin{eqnarray}
\label{a31}
a(t)&=&\alpha e^{-i(\omega\pm \epsilon\frac{1}{2}\sqrt{\alpha^{*}\alpha})t}\,,\\
\label{a32}
b(t)&=&\mp
\frac{i}{2}\sqrt{\frac{\alpha^{3}}{\alpha^{*}}}e^{-i2(\omega\pm\frac{1}{2} \epsilon
\sqrt{\alpha^{*}\alpha})t}\,.
\end{eqnarray}
{\em Remark}.The first integral of the system (\ref{e1a})--(\ref{e2a}) is of the form
$\omega a^{*}(t)a(t)+2\omega b^{*}(t)b(t)=const$. The solutions of Eqs. (\ref{a31})--(\ref{a32})
 naturally implies that $a^{*}(t)a(t)=const$ and $b^{*}(t)b(t)=const$ (see also \cite{Drobny}).
\subsection{Another selected solutions}
The method presented also allow us to find a periodic solution if the number of
equations is lager than two, for example \cite{Bandilla}:
\begin{eqnarray}
\label{a34}
 \frac{da}{dt}&=&-i\omega_{a}a +\epsilon c b^{*}\,,\nonumber\\
 \frac{db}{dt}&=&-i\omega_{b}b +\epsilon c a^{*}\,,\nonumber\\
 \frac{dc}{dt}&=&-i\omega_{c}c -\epsilon a b\,,
\end{eqnarray}
where $\omega_{c}=\omega_{a}+\omega_{b}$.
Eqs. (\ref{a34}) are used to describe the so-called  parametric optical processes
\cite{perina}.
If the initial conditions
denoted by $a(0)=a$, $b(0)=b$ and $c(0)=c$ satisfy the
relation $c=\frac{\mp ia
 b}{\sqrt{a^{*}\alpha+b^{*}b}}$ then the periodic solution is given by
\begin{eqnarray}
\label{a35}
 a(t)&=&a e^{-i\left(\omega_{a} \pm \epsilon
 \frac{b^{*}b}{\sqrt{a^{*}a+b^{*}b}}\right)t}\,,
\nonumber\\
 b(t)&=&b e^{-i\left(\omega_{b} \pm \epsilon
 \frac{a^{*}a}{\sqrt{a^{*}a+b^{*}b}}\right)t}\,,
\nonumber\\
c(t)&=&\frac{\mp ia
 b}{\sqrt{a^{*}a+b^{*}b}}\,  e^{-i\left(\omega_{c}
 \pm \epsilon \sqrt{a^{*}a+b^{*}b}\right)t}\,.
\end{eqnarray}
The method presented is also useful if the nonlinearities in a dynamical system
are of different rank (for example, Kerr effect in the presence of
second-harmonic generation \cite{Sanchez}):
\begin{eqnarray}
\label{app2}
\frac{da}{dt}&=&-i\omega a +\epsilon a^{*}b -i\kappa a^{*}a^{2}\,,\\
\label{app3}
\frac{db}{dt}&=&-i2\omega b -\frac{1}{2}\epsilon a^{2}\,.
\end{eqnarray}
We obtain:
\begin{eqnarray}
\label{app4}
a(t)&=&\alpha e^{-i\Omega t}\,,\\
\label{app5}
b(t)&=&\left(i\frac{\kappa}{2\epsilon}\alpha^{2}\mp
  \frac{1}{2}i\sqrt{\frac{\kappa^{2}}{\epsilon^{2}}\alpha^{4}
+\frac{\alpha^{3}}{\alpha^{*}}}\,\,
  \right) e^{-i2\Omega t}\,,\\\nonumber
\Omega&=&\omega + \frac{1}{2}\left(\kappa \alpha^{*}\alpha \pm
  \sqrt{\kappa^{2}\alpha^{*2}\alpha^{2} +\epsilon^{2}\alpha^{*}\alpha}\,\right)\,,
\end{eqnarray}
where $\kappa=\epsilon \lambda$ ($\lambda$ is a numerical coefficient).
The  assumption $\kappa=\epsilon \lambda$ is necessary in the method presented.
For $\lambda=0$  we get the solutions (\ref{a31})-(\ref{a32}).

%%%%%%%%%%%%%%%%%%%%%%%%%%

\end{document}